\documentclass[fleqn,10pt]{SelfArx} 

\usepackage{lineno}
\modulolinenumbers[5]
\usepackage[utf8]{inputenc}
\usepackage[T1]{fontenc}
\usepackage[english]{babel} 
\usepackage{cite}
\usepackage{amsmath,amssymb,amsfonts,mathrsfs}
\usepackage{algorithmic}
\usepackage{graphicx}
\usepackage{array}
\usepackage[caption=false,font=footnotesize,labelfont=sf,textfont=sf]{subfig}
\usepackage{textcomp}
\usepackage{booktabs} 
\usepackage{multirow}
\usepackage{listings}
\newcolumntype{M}[1]{>{\centering\arraybackslash}m{#1}}
\newcolumntype{N}{@{}m{0pt}@{}}
\def\BibTeX{{\rm B\kern-.05em{\sc i\kern-.025em b}\kern-.08em
    T\kern-.1667em\lower.7ex\hbox{E}\kern-.125emX}}

\usepackage{refcount}
\usepackage{xspace}
\newcommand{\ie}{i.e.,\xspace}
\newcommand{\eg}{e.g.,\xspace}
\newcommand{\aka}{a.k.a.,\xspace}
\newcommand{\etc}{etc.\xspace}

\newcommand{\mn}[1]{MareNostrum{#1}\xspace}
\newcommand{\fpu}{FPU\_$\mu$Kernel\xspace}
\newcommand{\mem}{MEM\_$\mu$Kernel\xspace}
\xspaceaddexceptions{=}

\usepackage[acronym]{glossaries}
\newacronym{bsc}{BSC}{the Barcelona Supercomputing Center}
\newcommand{\bsc}{\gls{bsc}\xspace}
\newacronym{ipc}{IPC}{Instructions Per Cycle}
\newcommand{\ipc}{\gls{ipc}\xspace}
\newacronym{os}{OS}{Operating System}
\newcommand{\os}{\gls{os}\xspace}
\newacronym{ear}{EAR}{Energy Aware Runtime}
\newcommand{\ear}{\gls{ear}\xspace}
\newacronym{earl}{EARL}{the EAR Library}
\newcommand{\earl}{\gls{earl}\xspace}
\newacronym{fma}{FMA}{Fused-Multiply-Add}
\newcommand{\fma}{\gls{fma}\xspace}

\usepackage{tcolorbox}
\newtcolorbox{tabox}{colback=blue!10!white,colframe=blue!30!black,title=Takeaways:}

\definecolor{color1}{RGB}{0,0,90} 
\definecolor{color2}{RGB}{0,20,20} 
\definecolor{gray}{RGB}{0,0,90} 

\usepackage{hyperref} 

\hypersetup{
	hidelinks,
	colorlinks,
	breaklinks=true,
	urlcolor=color2,
	citecolor=color1,
	linkcolor=color1,
	bookmarksopen=false,
	pdftitle={Title},
	pdfauthor={Author},
}

\usepackage{url}

\usepackage{listings} 
\definecolor{col1}{HTML}{1E88E5}
\definecolor{col2}{HTML}{D81B60}
\definecolor{col3}{HTML}{43A047}
\definecolor{col4}{HTML}{F4511E}
\definecolor{col5}{HTML}{E259FF}

\usepackage{pifont}
\RequirePackage[normalem]{ulem}
\RequirePackage{color}\definecolor{RED}{rgb}{1,0,0}\definecolor{BLUE}{rgb}{0,0,1}

\providecommand{\DIFdel}[1]{{\protect\color{RED}\sout{#1}}}
\providecommand{\DIFdel}[1]{}

\JournalInfo{} 
\Archive{} 

\PaperTitle{Introducing \mn5: 
A European pre-exascale energy-efficient system designed
to serve a broad spectrum of scientific workloads}
\PaperTitleShort{Introducing \mn5: A European pre-exascale system}

\Authors{%
Fabio Banchelli\textsuperscript{1}*,
Marta Garcia-Gasulla\textsuperscript{1},
Filippo Mantovani\textsuperscript{1},
Joan Vinyals\textsuperscript{1},
Josep Pocurull\textsuperscript{1},
David Vicente\textsuperscript{1},
Beatriz Eguzkitza\textsuperscript{1},
Flavio C. C. Galeazzo\textsuperscript{2},
Mario C. Acosta\textsuperscript{1},
Sergi Girona\textsuperscript{1}%
} 
\affiliation{
  \textsuperscript{1}\textit{Barcelona Supercomputing Center Barcelona, Spain},
  \textsuperscript{2}\textit{HLRS - University of Stuttgart, Germany}
} 
\affiliation{*\textbf{Corresponding author}: fabio.banchelli@bsc.es} 

\newcommand{\keywordname}{Keywords} 

\Abstract{
\mn5 is a pre-exascale supercomputer at the Barcelona Supercomputing Center (BSC), part of the EuroHPC Joint Undertaking. With a peak performance of 314 petaflops, \mn5 features a hybrid architecture comprising Intel Sapphire Rapids CPUs, NVIDIA Hopper GPUs, and DDR5 and high-bandwidth memory (HBM), organized into four partitions optimized for diverse workloads.
This document evaluates \mn5 through micro-benchmarks (floating-point performance, memory bandwidth, interconnect throughput), HPC benchmarks (HPL and HPCG), and application studies using Alya, OpenFOAM, and IFS. It highlights \mn5’s scalability, efficiency, and energy performance, utilizing the EAR (Energy Aware Runtime) framework to assess power consumption and the effects of direct liquid cooling. Additionally, HBM and DDR5 configurations are compared to examine memory performance trade-offs.
Designed to complement standard technical documentation, this study provides insights to guide both new and experienced users in optimizing their workloads and maximizing \mn5’s computational capabilities.
}

\begin{document}

\maketitle 

\tableofcontents 

\thispagestyle{empty} 

\section{Introduction}\label{secIntro}

\mn5, hosted at the Barcelona Supercomputing Center (BSC), is a pre-exascale supercomputer within the EuroHPC Joint Undertaking. It is designed to support high-performance computing (HPC) workloads, offering a peak computational capacity of~$314$ petaflops (PFlop/s). \mn5 features a hybrid architecture that integrates different technologies provided by multiple vendors and assembled by Atos. The system is organized into four primary partitions optimized for general-purpose and accelerated workloads:
{\em General Purpose Partition (GPP)}, based on Intel Sapphire Rapids CPUs, with a small fraction of nodes featuring HBM memory technology;
{\em Accelerated Partition (ACC)}, based on Intel Sapphire Rapids CPUs coupled with NVIDIA Hopper GPUs;
{\em General Purpose - Next Generation Partition}, based on the NVIDIA Grace Superchip;
{\em Accelerated - Next Generation Partition}, which is still under definition.
This document provides a report on HPC benchmarks (HPL and HPCG) for both the GPP and ACC partitions, along with a detailed evaluation of the GPP partition using architectural micro-benchmarks and applications.

The General Purpose Partition employs Intel Sapphire Rapids CPUs and provides configurations with DDR5 memory and high-bandwidth memory (HBM), targeting applications that require substantial memory throughput. 
GPP earned the spot~$\#22$ in the Top500 list of June 2024~\cite{top500_gpp}.

The Accelerated Partition combines Sapphire Rapids CPUs with NVIDIA Hopper GPUs, interconnected via NVLink and PCIe Gen5. Two additional partitions, featuring next-generation CPU and GPU technologies, are planned for deployment in the near future to further extend \mn5’s capabilities.
ACC earned the spot~$\#8$ in the Top500 list of June 2024~\cite{top500_acc} and the spot~$\#15$ in the Green500.

The benchmarking detailed in this document is structured into three levels of increasing complexity:
{\em i)} Results of micro-benchmarking focus on metrics such as floating-point performance, memory bandwidth, and interconnect throughput.
{\em ii)} Summaries of HPC benchmarks, including HPL and HPCG, are presented.
{\em iii)} Scalability and efficiency studies are conducted using real-world scientific applications such as Alya for fluid dynamics, OpenFOAM for computational fluid mechanics, and IFS for weather and climate modeling.

\mn5 integrates advanced energy monitoring and management systems through the EAR (Energy Aware Runtime) framework, enabling precise power consumption monitoring at both the component and job levels. This study leverages EAR to evaluate the impact of direct liquid cooling technology on energy efficiency, particularly for the densely packed compute nodes of \mn5.

Additionally, \mn5 serves as a testbed for emerging memory technologies, with a detailed comparison of HBM and DDR5 configurations provided.

This evaluation aims to establish a reference for scientists using \mn5 to produce scientific results with complex computational codes. This document complements the technical documentation typically provided with new HPC clusters\footnote{\url{https://bsc.es/supportkc/docs/MareNostrum5/intro}}. It is intended to assist both novice users, offering insights into the system's behavior, and expert users, helping them anticipate the challenges and benefits of running their scientific codes on \mn5.

The document is organized as follows:
Section~\ref{secSystem} provides an overview of the system, describing the different partitions of \mn5 and the underlying hardware components and system software;
Section~\ref{secMicroBm} is dedicated to low-level micro-architectural benchmarks;
Section~\ref{secHpcBm} summarizes the performance and energy results measured when running the Top500 benchmarks, HPL and HPCG;
Section~\ref{secApps} reports about the HPC applications performances and scalability with runs up to thousand of compute nodes;
Section~\ref{secHbm} studies the impact on performance and power consumption of two different memory technologies, DDR5 and HBM, when used for running scientific applications;
Section~\ref{secConclusions} recollect comments and conclusions.

\section{System Overview}\label{secSystem}

MareNostrum~5 is a pre-exascale EuroHPC supercomputer hosted at the Barcelona Supercomputing Center (BSC). The system has a total peak computational power of~$314$~PFlop/s and is supplied by Bull SAS, combining Bull Sequana XH3000 and Lenovo ThinkSystem architectures. The system is organized into four partitions with different technical characteristics, collectively meeting the needs of any HPC user.

In the following sections, we introduce the overall infrastructure of the MareNostrum~5 supercomputer, with a focus on the first two partitions, GPP and ACC, which were deployed and made available to users in Q2 2024.

\subsection{General Purpose Partition (GPP)}

The General Purpose Partition (GPP) of the MareNostrum 5 supercomputer is composed of dual-socket compute nodes powered by Intel Sapphire Rapids 8480+ CPUs with 56~cores per socket at 2~GHz.
These compute nodes are organized into three configurations:

\begin{itemize}
\item \textbf{DDR Compute Nodes}: 6,192 nodes, each with 16 DIMMs of 16GB DDR5 memory, totaling 256~GB per node.
\item \textbf{DDR-HM Compute Nodes}: 216 nodes, each with 16 DIMMs of 64GB DDR5 memory, providing 1~TB per node.
\item \textbf{HBM Compute Nodes}: 72 nodes, equipped with 2 DIMMs of 16GB DDR5 and an additional 128~GB of HBM2 memory for high-bandwidth applications\footnote{The CPU model in HBM nodes is Intel Xeon CPU Max 9480. Visit \href{https://www.intel.com/content/www/us/en/products/compare.html?productIds=231746,232592}{Intel Ark} for more details.}.
\end{itemize}

\begin{figure}[!htbp]
  \centering
  \includegraphics[width=\columnwidth]{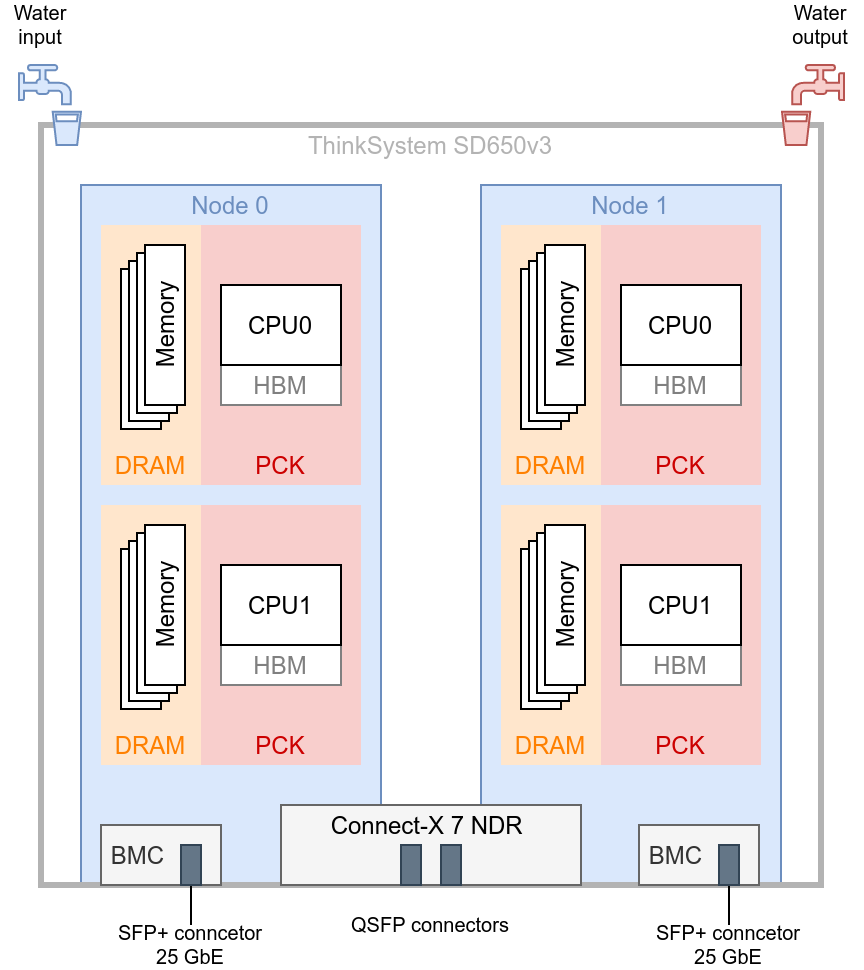}
  \caption[High level components connectivity of a \mn5 GPP tray which houses two compute nodes.]{High level components connectivity of a \mn5 GPP tray which houses two compute nodes\footnotemark[\getrefnumber{noteHBM}].}
  \label{figGppConnectivity}
\end{figure}

Each pair of compute nodes is installed on a ThinkSystem SD650v3 dual-node tray within a 19-inch chassis. These nodes share a single Infiniband NDR200 link, achieving up to 100 Gb/s bandwidth per node, alongside a dedicated 25~GbE link for service networking.
Figure~\ref{figGppConnectivity} shows the block diagram of the main components of a \mn5 GPP node\footnote{\label{noteHBM}HBM is depicted in gray because it is present only on the 216 nodes of the DDR-HM configuration.}.
The nodes are arranged in groups of 12 per chassis, which fit in a 6U space in each rack.

Cooling is handled with two circuits. One circuit, at 32-42°C, manages the direct liquid cooling of the compute nodes and power supplies ($2\times7.2$~kW), while the second circuit, at 17-27°C, cools the cold doors for air-cooled components. Power consumption varies by node type under HPL workloads (\ie 85\% of efficiency): 11.4~kW for DDR nodes, 12.0~kW for DDR-HM nodes, and 10.4~kW for HBM nodes.

Each rack contains, from bottom to top: three Lenovo chassis corresponding to 36 compute nodes, one Infiniband switch (model QM9790), another three Lenovo chassis corresponding to 36 nodes, and a top-of-rack Infiniband switch (model SN3700C).

\subsection{Accelerated Partition (ACC)}

\begin{figure}[!htbp]
  \centering
  \includegraphics[width=\columnwidth]{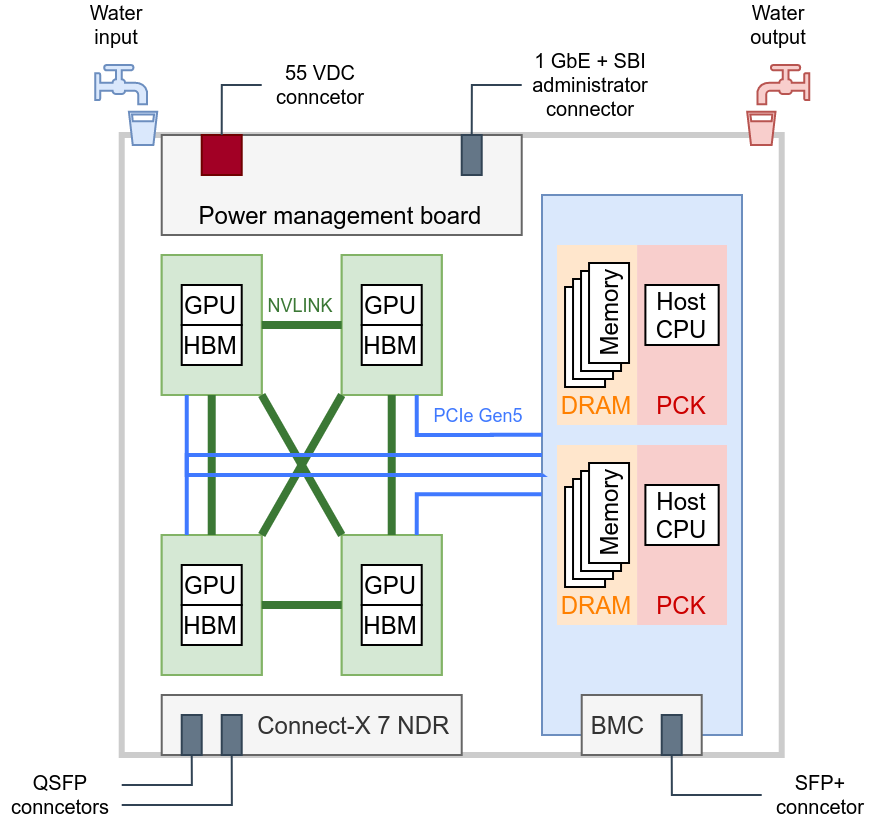}
  \caption{High level components connectivity of a \mn5 ACC compute node.}
  \label{figAccConnectivity}
\end{figure}

The compute nodes of the accelerated partition of \mn5 are composed of a host CPU board (blue in Figure~\ref{figAccConnectivity}) powered by two Intel Sapphire Rapids processors 8460Y+ (40~cores, 2.3~GHz, 300~W) and four NVIDIA Hopper GPU modules.
The CPU host board is equipped with 16 DIMMs of 32~GB running at 4.8~GHz DDR5, summing up 512~GB of RAM per ACC compute node.
Each Hopper GPU (H100) offers 64~GB of HBM2e and they are fully connected with NVLINK (green lines in Figure~\ref{figAccConnectivity}).
GPU pairs are connected with a bi-directional bandwidth of 300 GB/s.
Each of the four GPU modules provides a x16 PCIe Gen5 link for the upstream connectivity to the host CPU (blue lines in Figure~\ref{figAccConnectivity}).
Each node houses 480~GB NVMe local storage.

The compute node has been designed by Atos R\&D labs in France primarily for the \mn5 project.
All components of the ACC node are direct liquid cooled with the same hot-water circuit described for the General Purpose Partition.
Each GPU has a direct 200~Gb/s connection to the Infiniband network (ConnectX-7).
ACC compute nodes are interconnected with $2\times$ QSFP connectors, each one routing an Infiniband NDR200 link.
Power consumption under HPL workloads (\ie 77\% of efficiency) is 3.5~kW.

\subsection{Storage}
Each GPP node in \mn5 houses one NVME drive as local storage (omitted in Figure~\ref{figGppConnectivity} for simplicity).
This local storage can be configured upon request to provide users with a fast IO setup.
Beyond the node,
the storage infrastructure of \mn5 is divided into: HPC storage, archive storage and the parallel filesystem.

\paragraph{HPC storage} The building block of the HPC storage is the 5th generation of the Elastic Storage Server (model ESS-3500). 
Each ESS-3500 is powered by 
1x48 core AMD EPYC Rome running at 2.2~GHz,
512~GB of DDR4 memory,
$2\times$ NDR~200,
$2\times$ 100~Gbps ethernet,
over PCI Gen 4 links.
Depending on the kind of storage task to be performed, the ESS-3500 server can be configured as 
{\em i)} ESS-3500-P (as performance), \ie coupled with $24\times$ NVMe PCIe attached flash drives delivering 368~TB raw storage in 2U (net capacity~$2.81$~PB, $8+2$P);
{\em ii)} ESS-3500-C (as capacity), \ie coupled with $102\times$ HDD of 18~TB each HDD, delivering a total of 1836~TB raw storage in 4U (net capacity~$248$~PB, $8+3$P).
A storage unit ($\sim$ half rack) includes $2\times$~ESS-3500-P coupled with $4\times$~ESS-3500-C for a total of 18U and $\sim8$~PB of raw storage.
Combining these storage units into the 25 cabinet installed in \mn5 it allows a total raw capacity of 
4.79~PB of level~1 (or fast) storage (delivered by $13\times$~ESS-3500-P), combined with
367~PB of level~2 (or slow) storage (delivered by $50\times$~ESS-3500-C) and 4.79~PB.
Level~1 storage is logically dedicated to store data of highly accessed partitions, such as {\tt /home} or {\tt /apps} while level~2 storage is used for mid-term storage locations such as {\tt /projects} or {\tt /scratch}.

\paragraph{Archive storage}
The long term storage is handled with a tape system that includes $2\times$ IBM TS4500.
The archive of \mn5 offers a total of 44~PB disk cache combined with 400~PB total tape capacity.
The underling resources are~64 TS1160 fibre channel drives coupled with $20100\times$ JE Enterprise Gen6 tapes of 20~TB each, orchestrated by $8\times$ Spectrum Archive servers.

\paragraph{Parallel filesystem} The parallel filesystem adopted in \mn5 is GPFS. It is a high-performance, distributed filesystem optimized for scalability and fault tolerance in large computing clusters. Key features include:  
\begin{itemize}
\item POSIX-Compliant Access -- Supports standard system calls, allowing concurrent access to files by multiple processes across nodes.  
\item Data Striping -- Distributes file blocks across multiple disks in a round-robin fashion, improving aggregate bandwidth and load balancing.
\item Distributed Metadata -- Spreads metadata across nodes to avoid bottlenecks and enhance performance.  
\item Fault Tolerance -- Eliminates single points of failure and ensures data integrity despite node or disk failures.  
\item Scalability -- The parallel filesystem is able to sustain~$1.2$~TB/s and~$1.6$~TB/s of read and write bandwidth, respectively.  
\end{itemize}
GPFS is ideal for high-performance computing and big data environments, providing efficient, concurrent data access with robust fault tolerance.  

\subsection{Network}\label{secNetwork}

\begin{figure*}[!htbp]
  \centering
  \includegraphics[width=\textwidth]{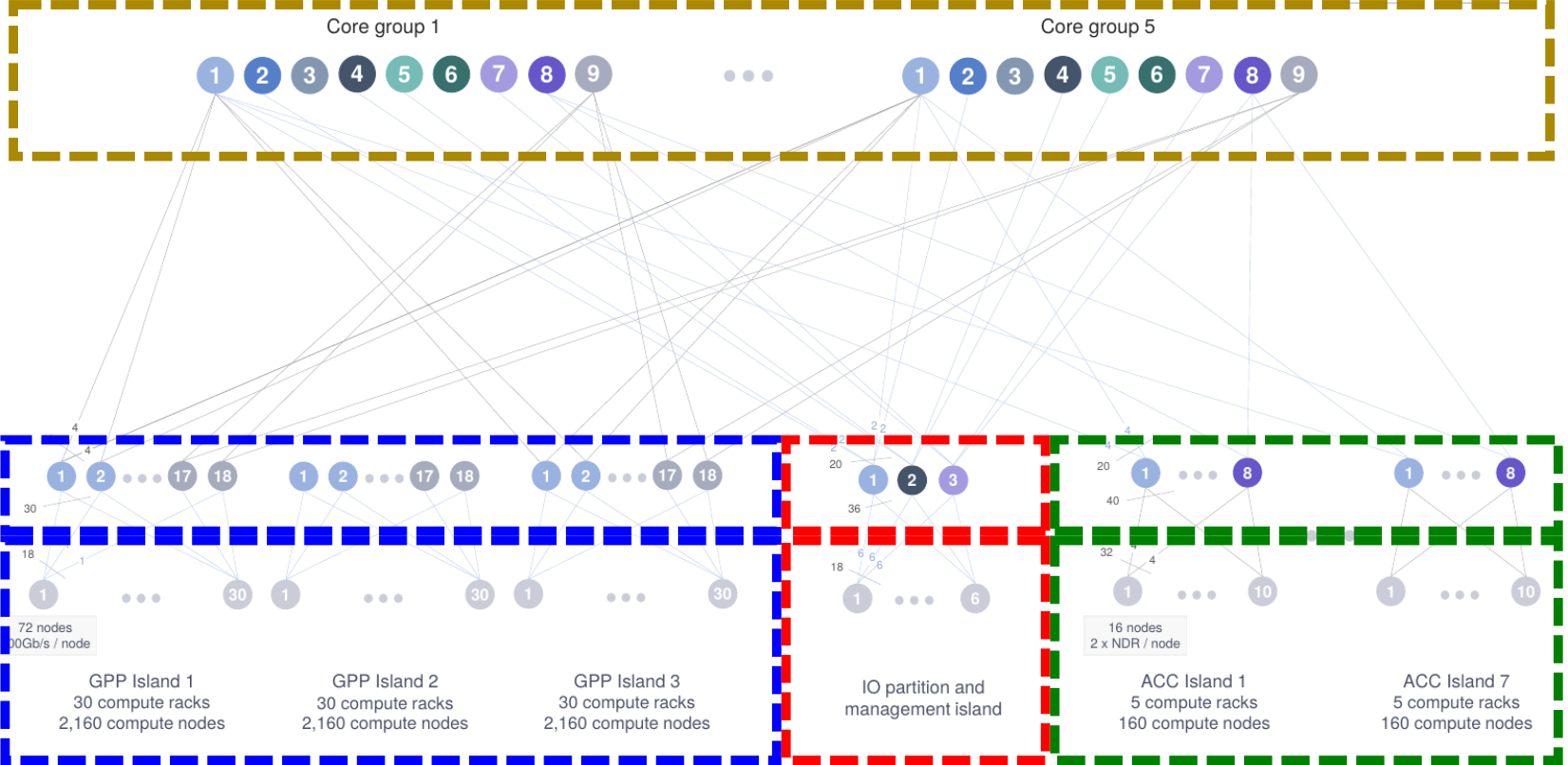}
  \caption{Network diagram of \mn5.}
  \label{figSystemNetwork}
\end{figure*}

\mn5 has a high-speed network that uses Infiniband NDR200 fabric.
The network topology is a three-layer fat-tree with a total of 324 switches (model QM9790). 
Figure~\ref{figSystemNetwork} represents a schematic view of the \mn5 network.
It has an implementation of 3 GPP islands (blue), 1 storage island (red) and 7 ACC islands (green). The layer levels are as follows:
    Layer 1 includes compute and storage nodes.
    Layer 2 connects the switches of the Layer 1 to the core switch.
    Layer 3 is the top-level layer including core switches (yellow).
For clarity, Figure~\ref{figSystemNetwork} does not depict all the connections between switches and nodes.

\subsection{System software}\label{secSoftware}

\paragraph{Operating System}
The OS in all compute nodes of \mn5 is an instance of Red Hat Enterprise Linux version~$9.2$ with a Linux kernel version~$5.14$.

\paragraph{Lmod}
Software tools and libraries are provided to users via environment modules with Lmod~\cite{lmod}.
Modules are classified into different types such as compilers, libraries, and applications.
Each module describes the dependencies and conflicts with other modules to ensure that users are able to configure a functioning software environment.
The list of available software changes over time, but it is always updated on the BSC website~\footnote{\url{https://www.bsc.es/marenostrum/marenostrum/available-software}}.

\paragraph{SLURM}
Access to compute resources is granted via the SLURM scheduler~\cite{yoo2003slurm}.
The SLURM configuration also includes job statistics to evaluate the current status of the cluster and perform statistical analysis of its usage.
\mn5 users may run interactive and batch jobs targeting one of the multiple partitions of the cluster.
Users may also toggle certain features such as access to hardware counters~(enabling \texttt{perfparanoid}), configuring a specific CPU frequency governor, or requesting exclusive access to the allocated nodes.

\paragraph{Compilers}
In this work we use a subset of the available compilers in \mn5.
These are the Intel compiler version~$2023.2$ with a new backend which is based on LLVM (\ie \texttt{icx}), and the GNU compiler~$13.2.0$.

\paragraph{MPI}
Similarly, in this work we use only two MPI flavors of all the ones available in \mn5.
These are Intel MPI and OpenMPI.
Throughout the evaluation, we ensure that the correct compiler backend is chosen by setting the appropriate environment variables.
For example, \texttt{OMPI\_CC} to specify the C compiler in OpenMPI.

\paragraph{PAPI}
PAPI~\cite{mucci1999papi} is a performance analysis tool that allows user code to read hardware counters.
It abstracts the implementation details of each CPU and provides a unified interface that is portable across architectures.
In this work we use PAPI version~$7.0$ to read the cycles and instructions counters of the Sapphire Rapids CPU.
The values reported by these counters are the base on which we conduct performance analysis in Section~\ref{secMicroBm} and Section~\ref{secApps}.

\paragraph{TALP}
TALP~\cite{talp} is the profiling tool that we use to gather the efficiency metrics presented in Section~\ref{secMethodology} (detailed definition can be found in Appendix~\ref{secMetrics}).
The current version of TALP supports MPI, OpenMP and hardware PAPI counters.
TALP can be used straightforward without any code modification nor recompilation of the code, it will automatically hook into the MPI and OpenMP calls and will use PAPI underneath to obtain cycles and instructions accounting.

TALP also provides an API to manually annotate the code, allowing to obtain the efficiency metrics for different code regions annotated by the user. 
One of the codes studied in the evaluation leverages this feature allowing a more detailed analysis of its performance.

\subsection{Power monitoring}\label{secPowerMonitoring}

\paragraph{EAR}
\ear~\cite{ear, ear_intel} is a tool that monitors power consumption and CPU usage in \mn5.
It can also apply CPU frequency scaling depending on its measurements although we do not include this feature in our work.

The measurement methodology of EAR varies depending on the monitoring infrastructure provided by the hardware vendor (\etc, RAPL counters, out-of-band power monitors, \etc).
In the case of the general purpose nodes in \mn5, EAR collects energy measurements of three components which are color-coded in Figure~\ref{figGppConnectivity}:
light-red each Intel CPU (\aka PCK or package),
light-orange each memory block (\aka DRAM),
and light-blue the whole node (\aka DC).
Measurements of PCK and DRAM are performed via RAPL counters while full-node power consumption is measured through an IPMI interface that queries a board management control unit.
A daemon process (\texttt{eard}) communicates with the respective hardware component that performs the measurements and stores the data in a database.
Even if the monitoring infrastructure in \mn5 provides energy measurements, \texttt{eard} averages power samples over a certain window of time and stores it as a timestamped series in a database.
Moreover, PCK and DRAM measurements are performed on each socket independently, but the \ear database stores the aggregated measurements of both sockets.

There are three agents that request measurements to \texttt{eard},
each one with a different sampling rate and serving a specific purpose.
The rest of this section is dedicated to explaining each one of these agents and a brief experiment to approximate the idle power of one CPU.

\noindent{\em 1. Periodic metrics} -- 
The first agent is implemented as a process that performs periodic measurements every minute and is always enabled.
The set of data collected by this agent is labeled {\em periodic metrics} and it is used by system administrators to observe the health of a node throughout its lifetime.
By default, this data is not available to users and it has no notion of SLURM jobs: the reader can see this as a coarse-grained free-running polling.

\noindent{\em 2. Job accounting} --
The second agent is triggered by a SLURM plugin that enables it at the beginning and disables it the end of a SLURM job step.
These data are also measured periodically, once every minute and stored into de EAR database.
Users can query the power and energy measurements of their jobs after they have ended.
Data is presented either by a time series that includes each sample, or as a summary of all measurements which reports the total energy and average power consumption.
We use this method to measure power consumption of scientific applications (Section~\ref{secApps}).

\noindent{\em 3. EAR Library} -- 
The third agent is disabled by default and can be enabled by users with the use of a SLURM flag.
The flag enables instrumentation by \earl, which spawns a thread that performs measurements every ten seconds ($6$~samples/min).
The data collected by this agent can be either stored in the \mn5 database or into a local database managed by the user.
We use this method to measure power consumption of micro-benchmarks in Section~\ref{secMicroBm}.

\begin{tabox}
  \begin{itemize}[leftmargin=*]
    \item PCK power includes CPU and HBM memory, if present
    \item DRAM power includes consumption of the DDR
    \item PCK and DRAM power aggregate both sockets
  \end{itemize}
\end{tabox}


\section{Low-level Benchmarks}\label{secMicroBm}

In this section we evaluate three major components of the GPP nodes in an isolated method.
We start by empirically measuring the idle CPU power consumption in GPP and then measure performance using micro-benchmarks.
In particular, we present the measurement of the performance of
the CPU by measuring the frequency and the performance in floating-point operations per cycle achieved,
we also show the bandwidth and latency of the memory subsystem and
the bandwidth and latency of the network.

\paragraph{Idle power}\label{secIdlePower}
We conduct an experiment to measure the power consumption while the node is idle.
Our test consists on allocating a node of \mn5 and running a SLURM job with one process executing the \texttt{sleep} command for~$60$s.
This puts one of the cores into sleep mode, while the other cores are considered to be idling.
We repeat this experiment on multiple nodes, always enabling \earl.
Since the power monitoring infrastructure exposed to users does not allow to read each CPU separately,
we assume that the idle power consumption of one CPU is half of the PCK measurement reported by \ear.
Similarly, we assume that the idle power consumption of the memory in one socket is half of the DRAM measurement.

\ear reports a PCK idle power of~$287.39$~W, DRAM of~$5.39$~W, and that the total node idle power is~$379.40$~W.
The standard deviation of PCK and node power is under~$6\%$ while the DRAM is~$18.59\%$.

\begin{tabox}
  \begin{itemize}[leftmargin=*]
    \item Idle power for one CPU is $143.70$~W
    \item Idle power for memory of one socket is $9.30$~W
    \item Idle power for one node is $379.40$~W
  \end{itemize}
\end{tabox}
\subsection{Floating-point unit}\label{secFpu}

\paragraph{Theoretical peak}
For any given floating point precision, let:
$V$ be the vector length of a SIMD unit, and
$N$ be the number of SIMD pipelines.
Assuming a constant throughput of one \fma instruction per cycle on each pipeline, we compute the theoretical peak performance $P$, in Flop/cycle as:
$$ P = V \times N \times 2 $$
Table~\ref{tabFpuTheoretical} lists the theoretical peak of \fma instructions in the Sapphire Rapids CPU for three ISAs and both single and double precision elements.
The naming convention of the first column is:
{\em isa}.{\em datatype},
where {\em isa} can be x86, avx2, or avx512;
and {\em datatype} can be sp, or dp.

\begin{table}[htbp]
  \centering
  \caption{Theoretical single-core performance of Sapphire Rapids}
  \label{tabFpuTheoretical}
  \begin{tabular}{lrrr}
  \multicolumn{1}{c}{\textbf{ISA}} &
  \multicolumn{1}{c}{\textbf{Vector Length}} &
  \multicolumn{1}{c}{\textbf{Pipelines}} &
  \multicolumn{1}{c}{\textbf{Performance}} \\
  &
  \multicolumn{1}{c}{$V$}   &
  \multicolumn{1}{c}{$N$}    &
  \multicolumn{1}{c}{$P$} \\ \midrule
  x86.sp    &  1 & 2 &  4 \\
  x86.dp    &  1 & 2 &  4 \\
  avx2.sp   &  8 & 2 & 32 \\
  avx2.dp   &  4 & 2 & 16 \\
  avx512.sp & 16 & 2 & 64 \\
  avx512.dp &  8 & 2 & 32
  \end{tabular}
\end{table}

The following experiments analyze if the CPU can sustain these theoretical performance numbers.

\paragraph{\fpu}
The \fpu is a synthetic benchmark developed at \bsc to measure the sustained floating-point performance of the CPU.
The source code is available upon request to the authors of this work.

The \fpu is written in C and parallelized with OpenMP.
The benchmark loops over a sequence of assembly instructions and measures their performance.
The type and number of floating point operations is known at compile time.
Runtime parameters allow to customize the amount of iterations to perform.

Performance is calculated based on three measurements:
{\em i)} cycles via PAPI (\texttt{PAPI\_TOT\_CYC}),
{\em ii)} instructions via PAPI (\texttt{PAPI\_TOT\_INS}), and
{\em iii)} micro-seconds via \texttt{gettimeofday}.
On top of the raw measurements, the benchmark reports 
performance (Flop/cycle and Flop/s),
frequency (MHz), and
\ipc.

For parallel runs, every thread performs the same amount of work (\ie weak scaling)
and threads synchronize at the end of each iteration.
Another runtime parameter allows to choose between reporting metrics per thread or summarized.
We combine the performance reported by the \fpu with the measurements performed by EAR to calculate the power efficiency of each type of instruction.

\paragraph{Single-core performance}
This experiment tries to measure the single-core performance of the Sapphire Rapids CPU and compare the measurements with the theoretical peak listed in Table~\ref{tabFpuTheoretical}.
In this experiment, we run the \fpu with different types of instructions and \earl is enabled.
We report the PCK power consumption measured by \ear subtracting the idle power of one CPU as explained in Section~\ref{secPowerMonitoring}.
%
%
With the data gathered with \earl we can also calculate the power efficiency of each type of instruction.
Table~\ref{tabFpuKernel} lists our results for each kernel.

\begin{table*}[htbp]
  \centering
  \caption{\fpu: single-core performance}
  \label{tabFpuKernel}
\resizebox{\textwidth}{!}{%
  \begin{tabular}{lrrrrrrrrrrrr}
  \multicolumn{1}{c}{\textbf{ISA}} &
  \multicolumn{1}{c}{\textbf{Instr.}} &
  \multicolumn{1}{c}{\textbf{Operations}} &
  \multicolumn{1}{c}{\textbf{Duration}} &
  \multicolumn{1}{c}{\textbf{Cycles}} &
  \multicolumn{1}{c}{\textbf{Freq.}} &
  \multicolumn{3}{c}{\textbf{Performance}} &
  \multicolumn{1}{c}{\textbf{IPC}} &
  \multicolumn{1}{c}{\textbf{Power}} &
  \multicolumn{1}{c}{\textbf{Efficiency}} \\
                                 &
  \multicolumn{1}{c}{$10^9$}     &
  \multicolumn{1}{c}{$10^7$}     &
  \multicolumn{1}{c}{s} &
  \multicolumn{1}{c}{$10^9$} &
  \multicolumn{1}{c}{GHz} &
  \multicolumn{1}{c}{Flop/Cycle} &
  \multicolumn{1}{c}{\%Peak}     &
  \multicolumn{1}{c}{GFlop/s}    &
                                 &
  \multicolumn{1}{c}{W} &
  \multicolumn{1}{c}{GFlop/(s$\times$W)} \\ \midrule
  x86.sp    & 1.93 &   3.20 & 2.69 &  8.00 & 2.98 &  4.00 & 100.00 &  11.90 & 2.41 & 152.70 & 0.08 \\
  x86.dp    & 1.93 &   3.20 & 2.70 &  8.00 & 2.97 &  4.00 & 100.00 &  11.87 & 2.41 & 158.94 & 0.07 \\
  avx2.sp   & 1.93 &  25.60 & 2.70 &  8.02 & 2.98 & 31.92 &  99.75 &  94.95 & 2.40 & 152.52 & 0.62 \\
  avx2.dp   & 1.93 &  12.80 & 2.70 &  8.02 & 2.98 & 15.96 &  99.75 &  47.49 & 2.40 & 159.46 & 0.30 \\
  avx512.sp & 3.53 & 102.00 & 7.30 & 16.00 & 2.20 & 63.80 &  99.69 & 140.37 & 2.20 & 152.35 & 0.92 \\
  avx512.dp & 3.53 &  51.20 & 6.49 & 16.00 & 2.47 & 31.91 &  99.72 &  78.86 & 2.20 & 152.43 & 0.52 
  \end{tabular}
}
\end{table*}

Our measurements show that the CPU frequency changes depending on the type of instruction being executed.
In the case of floating-point scalar and AVX2 code, frequency is consistently at $2.98$~GHz.
AVX512 instructions run at different frequencies depending on the precision:
$2.20$~GHz for avx512.sp, and
$2.47$~GHz for avx512.dp.
Automatic frequency adjustments depending on the type of instruction are a staple feature in modern Intel CPUs~\cite{avx512-freq}, but the specific frequency values and which instructions trigger them depend on the micro-architecture.

We observe that the sustained Flop/Cycle of each instruction type is within $1\%$~of the peak, meaning that the sustained performance matches the theoretical specs.
If the CPU could execute AVX512 instructions at $3.00$~GHz, it could theoretically achieve $94.00$~GFlop/s~($1.19\times$ the real sustained performance).

Looking at the power consumption, there is no real difference between any of the instruction types.
We observe an increase between~$9$~W and~$16$~W compared to the empirical idle power reported in Section~\ref{secIdlePower}.
All of our tests report a power consumption between~$152.35$~W and~$159.46$~W, which is a gap of less than~$5\%$.
It is interesting to note that avx512 instructions, even when running at a lower frequency, consume the same amount of power compared to other instruction types.

Regarding power efficiency, our results show that avx512.sp instructions are the most efficient ($0.92$~GFlop/(s$\times$W)) being~$50\%$ more efficient compared to the avx2.sp instructions.
On the other hand, the avx512.dp achieve slightly lower power efficiency compared to the avx2.dp instructions.

\begin{tabox}
  \begin{itemize}[leftmargin=*]
    \item Single-core performance matches theoretical peak
    \item There are three frequency levels that change depending on the type of instruction being executed
    \item avx512 is more power efficient than avx2
  \end{itemize}
\end{tabox}

\paragraph{Single-node performance}
This experiment analyzes the aggregated core performance and power efficiency of one node in \mn5.
We repeat the same executions as in the previous experiment with a full node, pinning one thread to each core, enabling \earl, and reporting the aggregated performance of all threads.
For this experiment, we take the total node power reported by \ear to calculate the power efficiency in a more representative metric of how much one node consumes.

\begin{figure}[!htbp]
  \centering
  \includegraphics[width=\columnwidth]{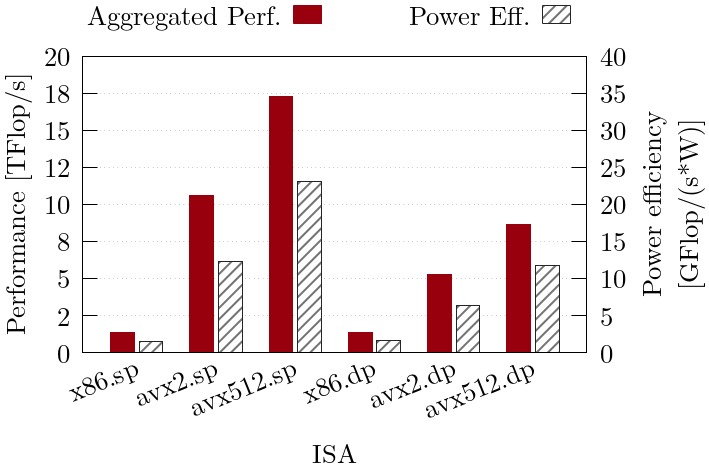}
  \caption{Single-node performance and power efficiency}
  \label{figFpuNode}
\end{figure}

Figure~\ref{figFpuNode} shows our results.
The left $y$-axis (solid red bars) represents performance in TFlop/s,
while right $y$-axis (dashed gray bars) represents power efficiency.

We observe that the aggregated performance is within~$5\%$ the theoretical peak even when all threads are running.
For avx512.dp the sustained performance is~$8.64$~TFlop/s.

Looking at the power efficiency, we observe that using wider SIMD instructions provides better power efficiency.
Comparing avx2.dp and avx512.dp, we see that the performance increases~$1.23\times$ at the cost of lowering the frequency~$10\%$ and consuming~$14\%$ less power.

During this experiment, we noticed that avx512 runs had at least one thread running slower than the rest.
We explore this effect in the following test.

\begin{tabox}
  \begin{itemize}[leftmargin=*]
    \item Aggregated performance of one node is $8.64$ GFlop/s
    \item Node power efficiency is $11.79$ GFlop/(s$\times$W)
  \end{itemize}
\end{tabox}

\begin{figure}[!htbp]
  \centering
  \includegraphics[width=\columnwidth]{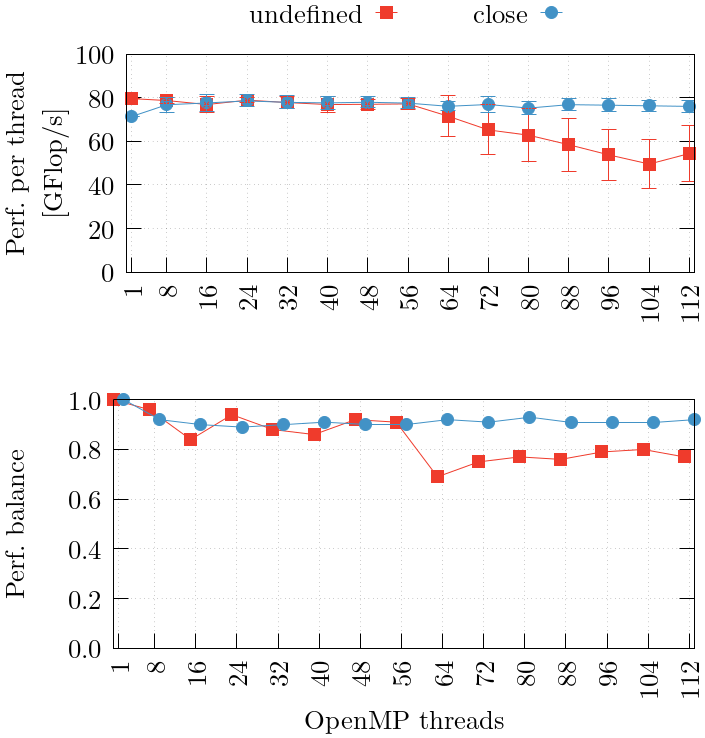}
  \caption{Average performance per thread, standard deviation, and performance balance.}
  \label{figFpuPinning}
\end{figure}

\paragraph{Performance balance and thread pinning}
This experiment tries to analyze the impact of \os noise by trying different thread pinning options.
We run the \fpu (avx512.dp) with increasing number of OpenMP threads and with three values of \texttt{OMP\_PROC\_BIND}:
undefined, \texttt{true}, and \texttt{close}.
All runs are done with \earl disabled.
Results are shown in Figure~\ref{figFpuPinning}.
Measurements with \texttt{OMP\_PROC\_BIND=true} are very similar to \texttt{close} and have been omitted for clarity.

The $x$-axis represents the number of threads.
The $y$-axis of the top plot represents average performance per thread in GFlop/s as reported by the benchmark, with the error bars representing the standard deviation.
The $y$-axis of the bottom plot represents the performance balance across all threads computed as the minimum performance across all threads divided by the average.
This metric has values between~$0$ and~$1$, and quantifies how badly the slowest thread performs compared to the average (higher is better).

We observe that the average performance per thread consistently stays around~$77$ GFlop/s when using \texttt{close}. Moreover, the standard deviation is small, always below~$5\%$ of the average.
However, the run with one thread performs remarkably low, achieving~$70.44$ GFlop/s.
This is probably due to a clash with other processes running on the same core, since the thread has been pinned to core~0.
This issue disappears when \texttt{OMP\_PROC\_BIND} is not defined, since the thread can be mapped to a different core and avoid sharing hardware resources with other processes.
When \texttt{OMP\_PROC\_BIND} is not defined, the average performance per thread drops for runs above~56 threads (one socket).
Our hypothesis is that there are thread migrations across sockets, but we have not conducted any further analysis to confirm that this is the case.

Looking at the performance balance, all runs with more than one thread have a balance of~$0.9$ or below (\ie the slowest thread runs~$10\%$ slower compared to the average).
For runs with small standard deviations, this means that all of the variability comes from one thread.
%
The single-thread run and \texttt{OMP\_PROC\_BIND} not defined does not suffer any performance loss.
However, the performance unbalance is not present when running the benchmark with other instruction types.
Our conclusion is that the \os impact comes from frequency scaling transitions since avx512 are the only instructions to have this variation.
We have also confirmed that this behavior is consistent across multiple nodes.

\begin{tabox}
  \begin{itemize}[leftmargin=*]
    \item Pinning reduces performance variability across threads
    \item Some system daemons are pinned to core zero
  \end{itemize}
\end{tabox}

\subsection{Memory hierarchy}\label{secBandwidth}
In this section, we analyze the memory hierarchy of the DDR5-based nodes.
For a comparison between DDR5 and HBM, please refer to Section~\ref{secHbm}.

\paragraph{Theoretical peak}
The micro-architecture of the Sapphire Rapids core includes three load pipelines,
each one being able to serve one cache line per cycle.
Similarly, there are two store pipelines.
Given these hardware resources, the theoretical L1 peak throughput that the core can achieve is three and two instructions per cycle for read and write operations, respectively.
The bandwidth depends on the type of instruction, if the data being accessed is cache-aligned, \etc
For example, an avx2 load instruction loads~$32$ bytes.
With all three load pipelines, the peak bandwidth of avx2 load instructions would be~$96$~Bytes/cycle.
Assuming a constant CPU frequency of~$3.00$~GHz, it would be equivalent to~$288$~GB/s.

On the opposite end, the peak memory bandwidth that the DDR5 can provide running at~$4800$~MHz is~$307.20$~GB/s per socket,
or~$614.40$~GB/s per node.

\paragraph{\mem}
The \mem is a synthetic benchmark developed at \bsc to measure the sustained memory bandwidth of the CPU.
It is based on the \fpu and shares with it the same code structure and measurement methodology.
Source code is available upon request.

The \mem has different kernels that perform a given amount of read and write instructions with different ISAs to contiguous memory positions that may or may not be cache aligned.
In this case, the amount of bytes read and written is known at compile time.
Bandwidth is reported as~Byte/cycle and~GB/s.
We use \mem to measure the bandwidth of the L1 cache in the Sapphire Rapids CPU.

\paragraph{L1 cache bandwidth}
In this experiment we aim to measure the L1 cache bandwidth using the \mem.
We use kernels that perform only read or only write operations to the L1 cache and report the sustained bandwidth.
Table~\ref{tabMemKernel} lists our results.

\begin{table*}[htbp]
  \centering
  \caption{\mem: single-core performance}
  \label{tabMemKernel}
  \begin{tabular}{lrrrrrrrrr}
  \multicolumn{1}{c}{\textbf{ISA}} &
  \multicolumn{1}{c}{\textbf{Instr.}} &
  \multicolumn{1}{c}{\textbf{Read}} &
  \multicolumn{1}{c}{\textbf{Written}} &
  \multicolumn{1}{c}{\textbf{Duration}} &
  \multicolumn{1}{c}{\textbf{Cycles}} &
  \multicolumn{1}{c}{\textbf{Freq.}} &
  \multicolumn{2}{c}{\textbf{Bandwidth}} &
  \multicolumn{1}{c}{\textbf{IPC}} \\
                                 &
  \multicolumn{1}{c}{$10^9$}     &
  \multicolumn{1}{c}{GB}         &
  \multicolumn{1}{c}{GB}         &
  \multicolumn{1}{c}{s}          &
  \multicolumn{1}{c}{$10^9$}     &
  \multicolumn{1}{c}{GHz}        &
  \multicolumn{1}{c}{Byte/Cycle} &
  \multicolumn{1}{c}{GB/s}       &
                                 \\ \midrule
  x86.rd    &  53.63 &  1280 &     0 & 17.99 & 160.31 & 2.98 &  23.87 &  71.14 & 2.99 \\
  avx2.rd   &  54.04 &  5120 &     0 & 18.13 & 160.31 & 2.98 &  94.75 & 282.38 & 2.97 \\
  avx512.rd &  84.87 & 10240 &     0 & 28.48 & 160.31 & 2.98 & 120.65 & 359.54 & 1.89 \\
  x86.wr    & 160.31 &     0 &  1280 & 26.86 &  80.08 & 2.98 &  15.98 &  47.65 & 2.00 \\
  avx2.wr   & 160.31 &     0 &  5120 & 26.86 &  80.08 & 2.98 &  63.94 & 190.59 & 2.00 \\
  avx512.wr & 160.31 &     0 & 10240 & 53.71 & 160.31 & 2.98 &  63.96 & 190.65 & 1.00
  \end{tabular}
\end{table*}

Firstly, we note that frequency stays at~$3.00$~GHz regardless of the type of instruction being executed.
Memory operations do not utilize floating-point functional units, so it is reasonable to conclude that they are not subject to the same power consumption limits as arithmetic operations.

Secondly, we observe that both x86 and avx2 instructions achieve an \ipc of~$3$.
This is somewhat expected, since the Sapphire Rapids CPU includes three load pipelines
and control instructions such as loop conditions and jumps have a negligible impact thanks to aggressive loop unrolling.
Our measurements show that such pipelines can handle from one to~$32$ double-precision elements per cycle.

On the other hand, the avx512 achieve an \ipc of~$2$.
This means that avx512 instructions cannot fully leverage the three load pipelines.
The reason behind this limitation comes from how many cache lines can be served from the L1 at any given cycle, which our measurements suggest that it is two lines per cycle.
Nonetheless, the sustained bandwidth of the avx512 is~$1.35\times$ better than the one of avx2.

In the case of store instructions, the scalar x86 instructions are able to leverage the two store pipelines,
but avx2 and avx512 instructions do not.
This is again due to the limitation of writing only one L1 cache line per cycle ($64$~Bytes/cycle).

\begin{tabox}
  \begin{itemize}[leftmargin=*]
    \item Single-core sustained read bandwidth is $359.54$~GB/s
    \item Single-core sustained write bandwidth is $190.65$~GB/s
  \end{itemize}
\end{tabox}

\begin{figure}[!htbp]
  \centering
  \includegraphics[width=\columnwidth]{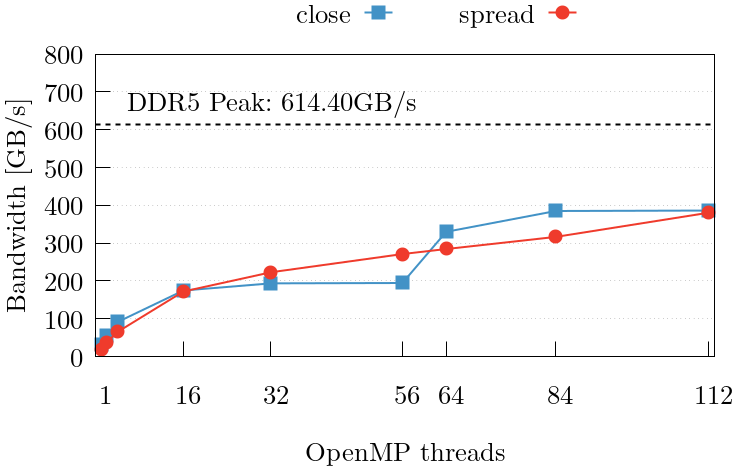}
  \caption{Aggregated copy bandwidth reported by \mem using two binding policies.}
  \label{figMemSingleNode}
\end{figure}

\paragraph{Single-node memory bandwidth}
In this experiment we use the \mem to measure the aggregated memory bandwidth across all threads when copying data.
Each thread allocates two arrays of~$64$~MiB, which is large enough to exceed the capacity of any cache in the hierarchy.
The kernel under study is written in C and we let the compiler auto-vectorize the code.
We use the optimization flag \texttt{-qopt-zmm-usage=high} to encourage the compiler to introduce avx512 instructions.
We also run the benchmark trying two values of \texttt{OMP\_PROC\_BIND}: \texttt{close} and \texttt{spread}.
Figure~\ref{figMemSingleNode} shows
the sustained bandwidth on the $y$-axis, and
the number of threads on the $x$-axis.

We observe that the \texttt{close} binding policy results on a curve with two steps.
This is expected, since executions up to~$56$ threads have all threads compete for memory resources on the same socket.
It is interesting to note that the sustained bandwidth within one socket flattens between~$16$ and~$32$ threads.
Adding the other socket brings more memory bandwidth to the table, but it again flattens starting at~$84$ cores, which corresponds with~$75\%$ of the cores in one node.

On the other hand, the \texttt{spread} binding policy allows threads to distribute evenly across the whole node and leverage all of the available bandwidth.
The curve does not have the same rough steps as with \texttt{close} and rises smoothly as the number of threads increases.
Up until~56 threads, the \texttt{spread} policy achieves better bandwidth since threads can leverage both sockets.
There is a window between~64 threads and the full node in which the \texttt{close} policy achieves higher bandwidth.
We repeated the test multiple times and observed the same results.
Furthermore, a close inspection of the bandwidth per thread reveals that, in the \texttt{spread} policy, there are some threads sustaining~$25\%$ less memory bandwidth, which translates to a lower aggregate bandwidth.
Once the node is full, there is no difference between the two policies.
The sustained bandwidth of a single DDR node is~$386.25$~GB/s, which corresponds to~$62.87\%$ of the theoretical peak.

%

\begin{tabox}
  \begin{itemize}[leftmargin=*]
    \item Single-core sustained memory bandwidth is $31.50$~GB/s
    \item Single-node sustained memory bandwidth is $62.87$~GB/s ($62.87\%$ of the peak)
    \item $84$ threads saturate the memory bandwidth of one node ($75\%$ of all the cores)
  \end{itemize}
\end{tabox}

\begin{figure}[!htbp]
  \centering
  \includegraphics[width=\columnwidth]{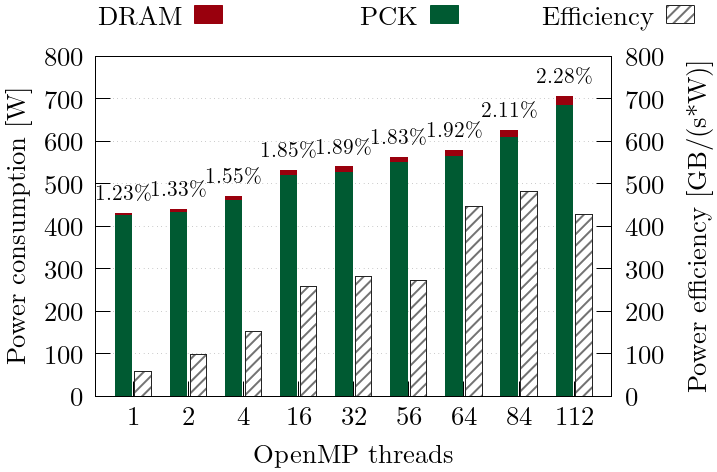}
  \caption{Power consumption and efficiency running \mem in \mn5 (\texttt{close} policy)}
  \label{figMemEfficiency}
\end{figure}

\paragraph{Power efficiency}
We take the aggregated memory bandwidth measured in the previous experiment using the \texttt{close} policy and combine it with the power measurements taken with \earl.
Figure~\ref{figMemEfficiency} shows the average DRAM and PCK power consumption when increasing the number of OpenMP threads.
The left $y$-axis represents the power in Watts, 
the right $y$-axis represents the power efficiency in~GB/(s$\times$W),
and while the labels on top of the bars indicate the percentage that the DRAM represents with respect to the whole node.

Our measurements show that, as we add more threads that perform memory operations, the impact of the DRAM on the overall power consumption of the node increases.
It starts with one thread at~$1.23\%$ and goes up to~$2.28\%$ with a full-node.
This is a very small portion of the total power consumption, but it highlights that the consumption of DRAM becomes more noticeable as more threads access it.
The reader should note that currently, we do not have a way to measure power consumption of caches in isolation.
This is included within the PCK measurement, which also increases with the number of threads.

Looking at the power efficiency, we see that it reaches a peak of~$481.29$~GB/(s$\times$W) when running with~84 threads.
After that point, the efficiency decreases.
This is another way to visualize how the memory bandwidth saturates when there are around~$75\%$ of the cores performing memory accesses.
After the saturation point, adding more threads increases the power consumption without giving any extra memory bandwidth.

\begin{tabox}
  \begin{itemize}[leftmargin=*]
    \item Power efficiency of the memory in one node is $481.29$~GB/(s$\times$W)
  \end{itemize}
\end{tabox}

\begin{figure}[!htbp]
  \centering
  \includegraphics[width=\columnwidth]{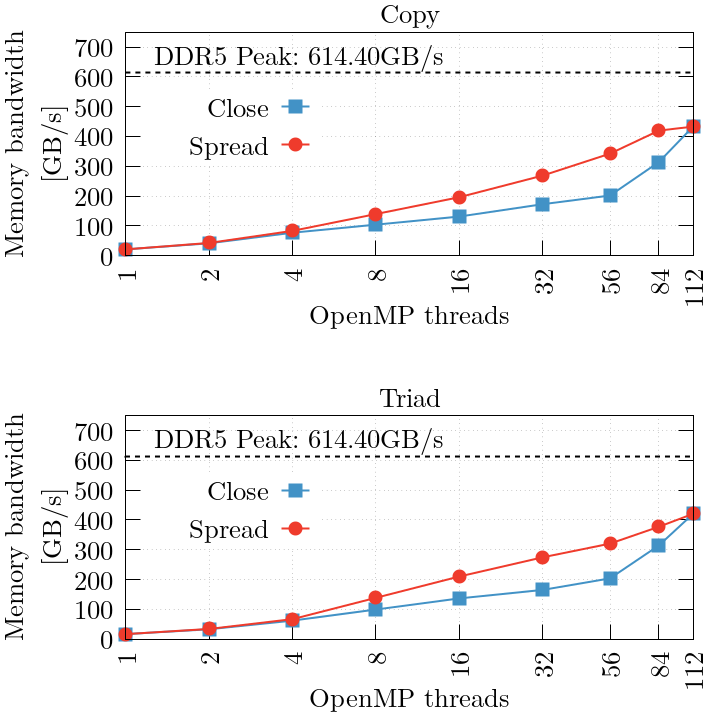}
  \caption{Sustained memory bandwidth running STREAM in \mn5 DDR nodes and using two binding policies. Dotted lines represent the theoretical peak.}
  \label{figStreamDdr}
\end{figure}

\paragraph{STREAM}
We also use the industry standard STREAM~\cite{mccalpin1995memory} benchmark (version 5.10) to measure the memory bandwidth in \mn5.
This code runs four kernels with varying arithmetic intensity: Copy, Scale, Add, and Triad.
Each kernel loops over an array of a fixed size known at compile time.
The rules of the benchmark state that the size of the arrays must be at least three times the size of the last level of cache, so for \mn5 we set a problem size of 83886080~elements (640~MiB per array).
This version of STREAM is parallelized with OpenMP, using worksharing clauses to implement strong scaling on each kernel.
We compile the code with the Intel compiler $2023.2.0$~and use the following optimization flags:
\texttt{-O3}, \texttt{-ffast-math}, and \texttt{-march=sapphirerapids}.
Figure~\ref{figStreamDdr} shows the memory bandwidth reported by STREAM in two relevant HPC kernels, Copy and Triad, when running in DDR nodes.
We also tried two mapping policies, with threads pinned to cores.

Compared to the results of the \mem, STREAM reports lower bandwidth with small pools of cores (\ie up to~$32$ threads),
but achieves higher bandwidth with high core counts.
We do not have an explanation for this discrepancies but we highlight two differences found by analyzing the binaries of each benchmark:
{\em i)} the compiler introduces software prefetching instructions in STREAM, but not in the \mem;
{\em ii)} STREAM opens and closes OpenMP parallel regions at each kernel execution, while \mem does it only once.

Looking at the shape of the bandwidth curves, we observe that STREAM shows the same trend as the \mem:
the spread mapping policy achieves better memory bandwidth when the node is not full.
This is expected since threads are evenly distributed across the two sockets of the node, balancing the memory requests to different memory controllers.

At best, our tests achieve $66.66\%$~of the theoretical peak with DDR nodes.
This result is comparable to measurements in older Intel-based CPUs such as Skylake~\cite{cte-cluster21,MANTOVANI2020800}.
The case of the Sapphire Rapids CPU is also discussed by McCalpin, the author of the STREAM benchmark, in~\cite{10.1007/978-3-031-40843-4_30}.


\paragraph{memlat}
memlat is a synthetic benchmark developed at \bsc to measure the single-core memory latency at different levels of the cache hierarchy.
This code performs a pointer chase through an array of a given size, $N$.
Each element of the array points to another element which is separated by a stride, $S$.
The benchmark performs a number of warm-up traversals $w$ and then runs for another $t$ traversals, which are timed.
$N$, $S$, $w$, and $t$ are runtime parameters.
The array allocation is aligned to a $4$~KiB page.
The code reports the measured average memory latency per memory access in micro-seconds and we convert this measurement to cycles assuming that the CPU frequency is~$3.00$~GHz.

\begin{figure}[!htbp]
  \centering
  \includegraphics[width=\columnwidth]{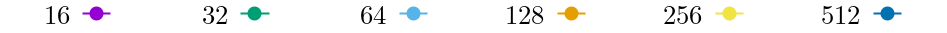}
  \includegraphics[width=\columnwidth]{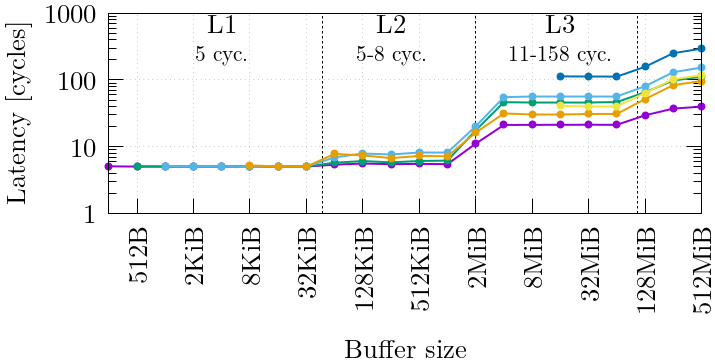}
  \caption{Measured latency of each level of the memory hierarchy in \mn5. Dotted lines represent the boundaries of each cache level. Different series correspond to different stride $S$ values.}
  \label{figMemlatCycles}
\end{figure}

\paragraph{Latency across the hierarchy}
Figure~\ref{figMemlatCycles} shows the measured latency with buffer of increasing size.
Each line corresponds to a different stride $S$ (64-bit elements).
The $y$-axis corresponds to latency in cycles.
Both axes are in logarithmic scale.
The plot is divided into four regions which correspond to each cache level and the main memory.
The image also includes the range of cycles measured for each region.

Accesses to L1 cache take~5 cycles consistently.
Accesses to L2 range from~5 to~8 cycles; and in L3 range from~11 to~158.
In the case of L2 and L3 accesses, the latency depends on the stride size.
The reader should consider that the effective latency for any given buffer size is within the range of the measured values for different values of $S$.
For small strides, the cache pre-fetcher is able to recognize the access pattern and reduce the measured latency.
On the contrary, bigger strides approximate better the actual memory latency.
Measurements in the L3 and beyond have higher variability and a wider range of latency because they are memory resources outside the core itself and shared across cores.

\begin{tabox}
  \begin{itemize}[leftmargin=*]
    \item Latency to the L1 cache is within~$5$ cycles
    \item Hardware prefetching makes it difficult to accurately measure latency to higher cache levels
    \item Depending on the amount of prefetching, accessing the L3 cache is between~$2\times$ and~$32\times$ slower than L1
  \end{itemize}
\end{tabox}

\subsection{Network bandwidth and latency}\label{secNetwork}
\paragraph{Network topology and theoretical peak}

In this section, we classify pairs of nodes in four categories:

\noindent{\em 1. Same tray} -- 
Messages reach the network controller, but do not leave the tray, so they are not limited by Infiniband.

\noindent{\em 2. Within the same rack} --
Messages go through level-one network switches, which are shared between~72 nodes.

\noindent{\em 3. Same island} --
Messages go through level-two network switches, which are shared between~30 racks~(2160 nodes).

\noindent{\em 4. Different islands} --
Messages go through level-three network switches, which there are nine in total and connect the whole general purpose partition of \mn5.

Please refer to Section~\ref{secNetwork} for a detailed description of the network connectivity in \mn5.
The theoretical peak bandwidth of the Infiniband NDR200 network fabric is~$25$~GB/s and the theoretical latency is under~$5$~us.
Pairs of nodes that share a tray may achieve better network speeds, but we could not find information on which is the theoretical peak.

\paragraph{OSU benchmarks}
We use the OSU benchmarks~7.4~\cite{osu-benchmarks} to evaluate the network capabilities of \mn5.
This is a benchmark suite that includes codes for different parallel programming paradigms and languages.
In this work, we focus on two of the simplest benchmarks, which are \texttt{osu\_bw} and \texttt{osu\_latency}.
Both codes are written in C and parallelized with MPI.
They perform multiple point-to-point messages between two processes and report the sustained bandwidth and latency, respectively.

\begin{figure}[!htbp]
  \centering
  \includegraphics[width=\columnwidth]{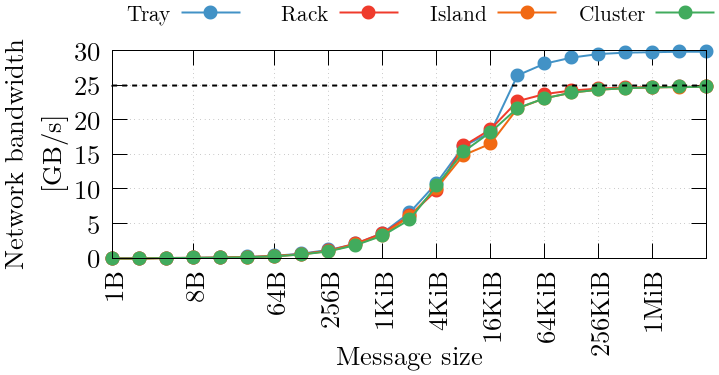}
  \caption{Network bandwidth between pairs of processes running \texttt{osu\_bw} in \mn5. The dotted line represents the theoretical peak of the Infiniband network.}
  \label{figOsuBw}
\end{figure}

\paragraph{Network bandwidth}
In this experiment, we run \texttt{osu\_bw} with OpenMPI~4.1.5 having pairs of MPI ranks mapped to different locations of the general purpose partition.
Processes are always in different nodes, but the distance between nodes varies.
Figure~\ref{figOsuBw} shows the bandwidth measured with \texttt{osu\_bw} with increasing message sizes.
The bandwidth reported by the benchmark increases with the message size.

We observe a breaking point around messages of~$4$~KiB, which corresponds to a transition in the communication protocol (rendezvous to eager).
For messages of~$64$~KiB and bigger, pairs of processes are able to match the theoretical peak of the Infiniband NDR200 ($25$~GB/s) and there is no meaningful difference between all but one of the lines.
In the case of processes within the same tray, our measurements show a greater bandwidth that reaches up to~$29.83$~GB/s.
As stated previously, this is due to the fact that these type of pairs do not require the Infiniband fabric to communicate.

\begin{figure}[!htbp]
  \centering
  \includegraphics[width=\columnwidth]{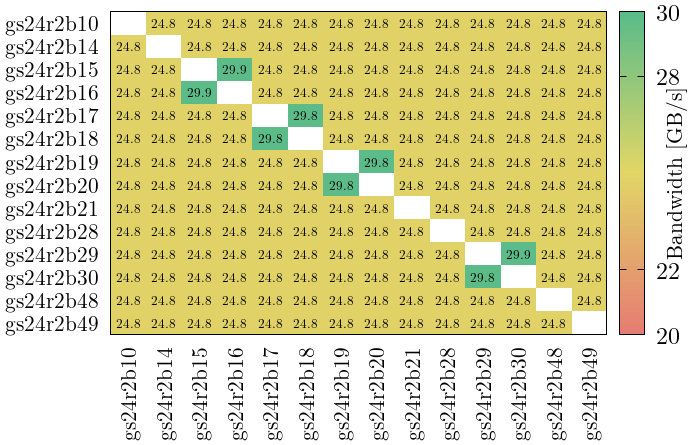}
  \caption{Network bandwidth between pairs of nodes in \mn5.}
  \label{figWeakLinks}
\end{figure}

\paragraph{Weak links}
In this experiment, we try to identify pairs of nodes which cannot sustain~$25$~GB/s.
We call these pairs "weak links".
To do so, we repeat the \texttt{osu\_bw} experiment with multiple pairs of nodes in \mn5.
Figure~\ref{figWeakLinks} shows a heatmap in which columns and rows represent nodes and each cell represents the network bandwidth achieved by a given pair of nodes.
Cells are color-coded and contain the value in~$GB/s$ reported by \texttt{osu\_bw}.

The reader should note that nodes in Figure~\ref{figWeakLinks} are not consecutive.
Given the amount of nodes in \mn5, it is unfeasible to check every possible pair, so we test only on a subset of~15 nodes, all of which are from the same island.

In line with the results from our previous experiment, pairs of nodes within the same tray reach up to~$30$~GB/s, while other pairs achieve around~$25$~GB/s.

\begin{figure}[!htbp]
  \centering
  \includegraphics[width=\columnwidth]{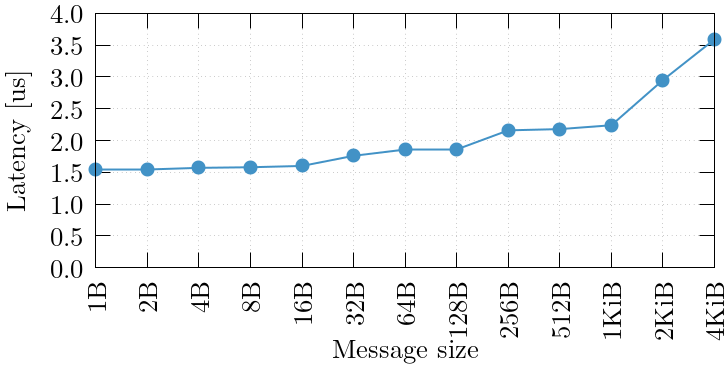}
  \caption{Network latency between pairs of processes within the same rack.}
  \label{figOsuLatency}
\end{figure}

\paragraph{Network latency}
In this experiment, we run \texttt{osu\_latency} with the same methodology as with \texttt{osu\_bw}.
This code measures the time that MPI messages take to go from one process to another, not the actual latency of the communication fabric.
Figure~\ref{figOsuLatency} shows the latency reported by \texttt{osu\_latency} when increasing the message size.
Processes are mapped to nodes in the same rack, but different trays.

We observe a minimum message latency of~$1.50$~us when the message size is under~$16$~B.
With such a small message, we can assume that this latency corresponds to the physical transport layer plus the overhead introduced by the MPI implementation.
As the message size increases, the latency reported by \texttt{osu\_latency} increases, since it requires exchanging more packets of data through the network.
In this experiment, we do not include messages bigger than~$4$~KiB because at that point, the communication will be bound by the bandwidth of the network, which we have already discussed in the previous experiments.

\begin{tabox}
  \begin{itemize}[leftmargin=*]
    \item Sustained point-to-point network bandwidth reaches theoretical peak ($25$~GB/s)
    \item Network latency for small messages is~$1.50$~us
  \end{itemize}
\end{tabox}

\section{HPC Benchmarks}\label{secHpcBm}

In this section we present the sustained performance of two major HPC benchmarks that are used to rank supercomputers in the Top500 list, HPL and HPCG.
Furthermore, we present the power consumption and efficiency of HPL.
We include single and multiple node results of both GPP and ACC partitions.
The reader should note that multi-node runs may not use all the available nodes in \mn5 because we present the results of runs with a parameter configuration (\eg \texttt{N}, \texttt{P}, \texttt{Q}) that yields the highest performance.

Moreover, we report the performance of widely used AI benchmarks (HPL-MxP, Graph500, and DL-BENCH) when running in the ACC partition.

\subsection{HPL}
The following paragraphs describe how we run HPL in \mn5 and Table~\ref{tabHPL} summarizes the performance and power efficiency results.

\paragraph{GPP}
In the case of the GPP partition, the HPL
binary was provided by Intel and includes MKL optimizations.
The execution maps~$2$ MPI processes per node (\ie one process per socket) with~$56$ OpenMP threads per process.
The execution with the full partition earned the spot~$\#22$ in the Top500 list of June 2024~\cite{top500_gpp} reaching~$40.10$~TFlop/s which accounts for~$92.19\%$ of the peak.

\paragraph{ACC}
In the case of the ACC partition, the HPL
binary was provided by Nvidia through Nvidia NGC as a Singularity image.
The execution maps~$4$ MPI processes per node (\ie one process per GPU) limiting the number of threads per GPU to~$20$.
The execution with the full partition earned the spot~$\#8$ in the Top500 list of June 2024~\cite{top500_acc} and the spot~$\#15$ in the Green500.

\begin{table}[htbp]
  \centering
  \caption{HPL single and multi node results in \mn5}
  \label{tabHPL}
\resizebox{\columnwidth}{!}{%
  \begin{tabular}{lrrrrrr}
  \multicolumn{1}{c}{\textbf{Partition}} &
  \multicolumn{1}{c}{\textbf{Nodes}}     &
  \multicolumn{1}{c}{\textbf{RPeak}}     &
  \multicolumn{1}{c}{\textbf{RMax}}      &
  \multicolumn{1}{c}{\textbf{\%Rpeak}}   &
  \multicolumn{1}{c}{\textbf{Power}}     &
  \multicolumn{1}{c}{\textbf{Efficiency}} \\
                                 &
                                 &
  \multicolumn{1}{c}{TFlop/s}    &
  \multicolumn{1}{c}{TFlop/s}    &
                                 &
  \multicolumn{1}{c}{kW}         &
  \multicolumn{1}{c}{GFlop/(s$\times$W)} \\ \midrule
  GPP & 1 &   7.17 &   6.61 & 92.19 & 0.87 &  7.59 \\
  ACC & 1 & 232.00 & 179.70 & 77.46 & 3.50 & 66.29 \\
  \end{tabular}
} \\ \vspace{2em}
\resizebox{\columnwidth}{!}{%
  \begin{tabular}{lrrrrrr}
  \multicolumn{1}{c}{\textbf{Partition}} &
  \multicolumn{1}{c}{\textbf{Nodes}}     &
  \multicolumn{1}{c}{\textbf{RPeak}}     &
  \multicolumn{1}{c}{\textbf{RMax}}      &
  \multicolumn{1}{c}{\textbf{\%Rpeak}}   &
  \multicolumn{1}{c}{\textbf{Power}}     &
  \multicolumn{1}{c}{\textbf{Efficiency}} \\
                                 &
                                 &
  \multicolumn{1}{c}{PFlop/s}    &
  \multicolumn{1}{c}{PFlop/s}    &
                                 &
  \multicolumn{1}{c}{MW}         &
  \multicolumn{1}{c}{GFlop/(s$\times$W)} \\ \midrule
  GPP & 6264 &  44.90 &  40.10 & 89.31 & 5.75 &  6.97 \\
  ACC & 1080 & 250.56 & 175.30 & 69.96 & 4.16 & 42.15 \\
  \end{tabular}
}
\end{table}

\subsection{HPCG}
The following paragraphs describe how we run HPCG in \mn5 and Table~\ref{tabHPCG} summarizes the performance results.

\paragraph{GPP}
In the case of the GPP partition, the HPCG
binary was provided by Intel and includes MKL optimizations.
The execution maps~$8$ MPI processes per node (\ie four processes per socket) with~$11$ OpenMP threads per process.
The execution with the full partition earned the spot~$\#26$ in the HPCG500 list of June 2024.

\paragraph{ACC}
In the case of the ACC partition, the HPL
binary was provided by Nvidia through Nvidia NGC as a Singularity image.
The MPI processes and threads per GPU are configured in the same way as with HPL:~$4$ MPI processes per node (\ie one process per GPU) limiting the number of threads per GPU to~$20$.
The execution with the full partition earned the spot~$\#12$ in the HPCG500 list of June 2024.

\begin{table}[htbp]
  \centering
  \caption{HPCG single and multi node results in \mn5}
  \label{tabHPCG}
\resizebox{.8\columnwidth}{!}{%
  \begin{tabular}{lrrrr}
  \multicolumn{1}{c}{\textbf{Partition}} &
  \multicolumn{1}{c}{\textbf{Nodes}}     &
  \multicolumn{1}{c}{\textbf{RPeak}}     &
  \multicolumn{1}{c}{\textbf{RMax}}      &
  \multicolumn{1}{c}{\textbf{\%Rpeak}}   \\
                                 &
                                 &
  \multicolumn{1}{c}{TFlop/s}    &
  \multicolumn{1}{c}{TFlop/s}    &
                                 \\ \midrule
  GPP & 1 &   7.17 & 0.08 & 1.16 \\
  ACC & 1 & 232.00 & 1.06 & 0.46 \\
  \end{tabular}
}
\\ \vspace{2em}
\resizebox{.8\columnwidth}{!}{%
  \begin{tabular}{lrrrr}
  \multicolumn{1}{c}{\textbf{Partition}} &
  \multicolumn{1}{c}{\textbf{Nodes}}     &
  \multicolumn{1}{c}{\textbf{RPeak}}     &
  \multicolumn{1}{c}{\textbf{RMax}}      &
  \multicolumn{1}{c}{\textbf{\%Rpeak}}   \\
                                 &
                                 &
  \multicolumn{1}{c}{TFlop/s}    &
  \multicolumn{1}{c}{TFlop/s}    &
                                 \\ \midrule
  GPP & 6270 &  44943.36 &  484.00 & 1.08 \\
  ACC & 1024 & 237568.00 & 1145.98 & 0.48 \\
  \end{tabular}
}
\end{table}

\subsection{HPL-MxP}
HPL-MxP is an HPC benchmark that is based on the traditional HPL benchmark but it includes mixed precision algorithms.
In recent years, this HPL-MxP has become an important benchmark to evaluate performance of HPC systems that target AI workloads.

We run a binary HPL-MxP provided by Nvidia in the ACC partition.
The execution maps~$4$ MPI processes per node and expands across~$1080$ nodes, achieving~$1.836$ EFlop/s.

\subsection{Graph500}
Graph500 is a benchmark that aims to complement traditional HPC benchmarks such as HPL and HPCG by introducing data intensive kernels that are used in graph-related codes.

A BFS run with~$1024$ ACC nodes, mapping~$4$ tasks per node and setting the scale to~$35$ scores a performance of~$15737.43$~GTEPS.
Similar to the Top500, there is also a Graph500 list.
The ACC partition of \mn5 has the earned the spot~$\#9$ on the November~$2024$ BFS list~\cite{graph500_acc}.

\subsection{DL-BENCH}
DL-BENCH consists of one single PyTorch code which is called \texttt{pytorch\_imagenet\_resnet50.py}, extracted directly from the horovod repository\footnote{\url{https://github.com/horovod/horovod/blob/v0.18.2/examples/}}.
It runs a distributed training across the tested system, using the popular resnet50 neural network to train on the Imagenet dataset, for 90 Epochs (\texttt{--epochs 90}).
Performance is reported as the elapsed time to complete the training.

We use this benchmark to evaluate the performance at a smaller scale,
with runs in the ACC partition using~$16$ and~$78$ nodes.
Table~\ref{tabDlBench} shows the results of DL-BENCH.

\begin{table}[htbp]
  \centering
  \caption{DL-BENCH results in the ACC partition}
  \label{tabDlBench}
\resizebox{.8\columnwidth}{!}{%
  \begin{tabular}{lrrrr}
  \multicolumn{1}{c}{\textbf{Partition}} &
  \multicolumn{1}{c}{\textbf{Nodes}}     &
  \multicolumn{1}{c}{\textbf{Time}}      &
  \multicolumn{1}{c}{\textbf{Energy}}    &
  \multicolumn{1}{c}{\textbf{Power}}    \\
                                 &
                                 &
  \multicolumn{1}{c}{s}          &
  \multicolumn{1}{c}{kWh}        &
  \multicolumn{1}{c}{kW}         \\ \midrule
  ACC & 16 & 2473.04 & 32.76 &  47.68 \\
  ACC & 78 & 644.254 & 33.36 & 186.41 \\
  \end{tabular}
}
\end{table}

\section{HPC Applications Scalability}\label{secApps}
After measuring the capabilities of the hardware resources of the cluster in isolation,
and the sustained performance achieved by standard HPC benchmarks,
in this section, we will evaluate the scalability of three scientific applications. 
The scientific applications chosen cover computational fluid dynamics (CFD) and numerical modeling or weather and climate prediction.
%
%
These two scientific domains represent a significant part of the computational resources of datacenters worldwide and have a recognized research impact~\cite{JANSSON2024106243,egusphere,banchelli2020benchmarking}.

All three applications are studied following the same methodology, which is described in Section~\ref{secMethodology}.
For a full breakdown of the software environment and compiler configuration used for each application,
contact the corresponding author of this work.


\subsection{Tools and methodology}
\label{secMethodology}

The methodology used to evaluate each one of the applications will be the same one and consists of three steps:
\begin{enumerate}
  \item Scalability: Measure elapsed time and throughput (note that for each application, a different throughput metric is defined, it should be meaningful for users and developers of the code) when scaling to an increasing number of nodes.
  \item Efficiency metrics: Obtain efficiency metrics for the scaling runs. The efficiency metrics used are formally described in Appendix~\ref{secMetrics} and are based on the POP methodology~\cite{pop_metrics}.
  \item Energy consumption: Gather the energy consumption and the Energy-Delay-Product for the scaling runs. 
\end{enumerate}

\paragraph{Focus of Analysis}
The region under study of each application is called focus of analysis.
All of the codes analyzed in this section have a similar execution structure with an initialization phase, an iterative execution phase, and a finalization phase.
Generally, the focus of analysis excludes the initialization and finalization phase.
It may also exclude some of the iterations of the execution phase.

The focus of analysis of each application studied in this work is explained at the beginning of its corresponding section.

\begin{figure}[!htbp]
  \centering
  \includegraphics[width=\columnwidth]{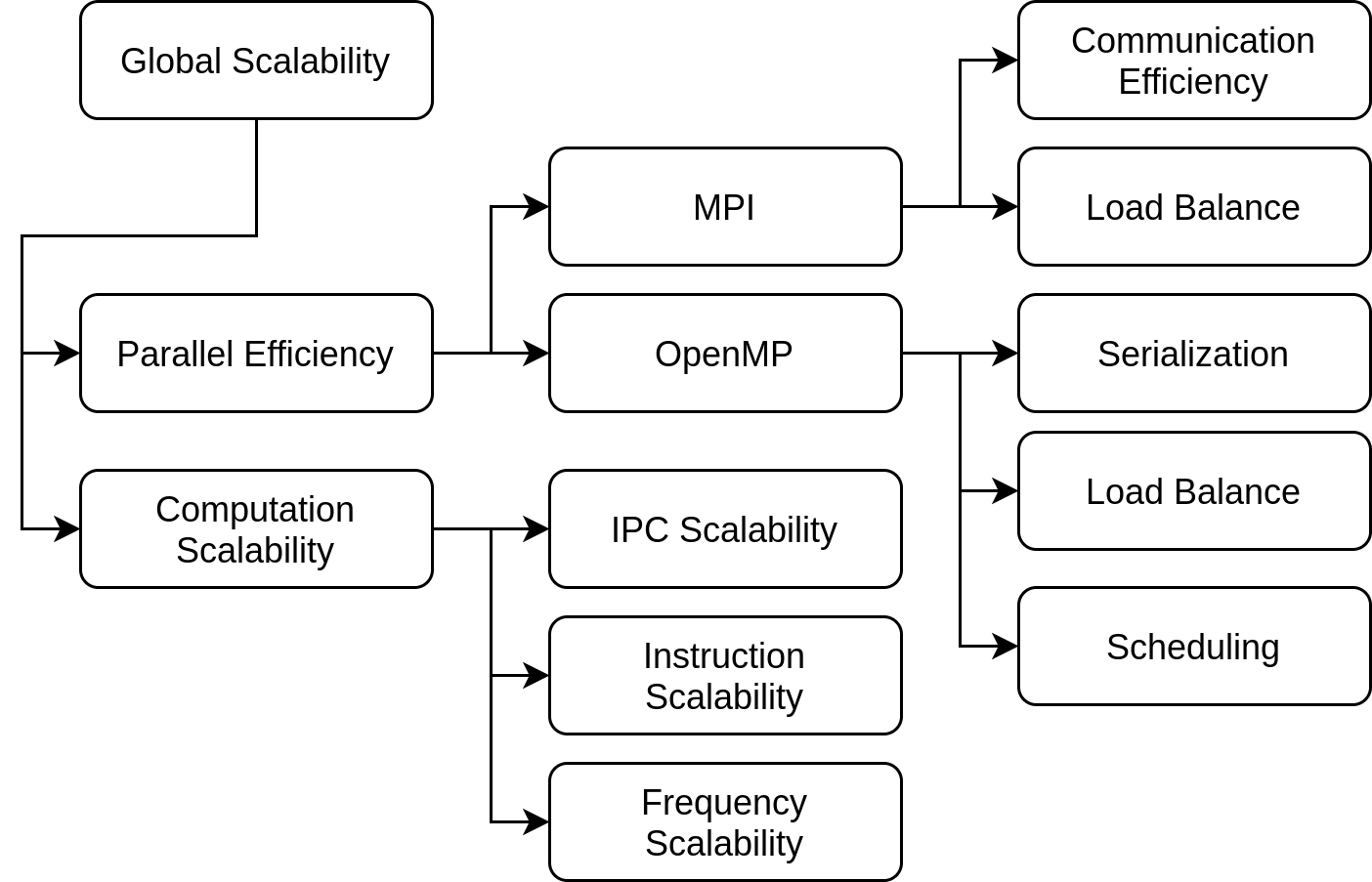}
  \caption{Efficiency metrics model.}
  \label{figAppsEfficiencyMetrics}
\end{figure}

\paragraph{Efficiency metrics model}
The Efficiency metrics model defines a collection of metrics arranged in a hierarchy.
Figure~\ref{figAppsEfficiencyMetrics} shows a schematic breakdown of the metrics displayed as a tree.
For all metrics, higher values are better.
Moreover, all metrics with the exception of the ones labeled {\em Scalability} take values between zero and one.

The {\em Global Scalability} is the root of the tree and it has two children metrics, {\em Parallel Efficiency} and {\em Computational Scalability}.
The former represents the efficiency of the parallel programming model and parallelization strategy of the code.
The later focuses on the change of useful computation across different executions.
Please refer to Appendix~\ref{secMetrics} for a thorough list of the metrics and their formulation.

\paragraph{Energy consumption}
Let $W$ be the power consumption of a whole node, in Watts, averaged across all nodes $N$ and throughout the whole execution.
Let $T$ be the elapsed time of the region under study of each application in seconds.
We compute the total energy consumption $E$, in $kWh$, as
$ E = W \times N \times T / 3600$.
And the Energy-Delay-Product $EDP$, in $kWh^2$, as
$ EDP = E \times T $.

This definition of $E$ and $EDP$ combines measurements in two overlapping, but not equal, windows of time.
The average power $W$ takes into account the whole duration of a job,
while the time $T$ takes into account only our focus of analysis.
For jobs in which there is a lot of time outside the focus of analysis in which the power consumption behaves completely different than inside, our definition may not accurately represent the energy consumption of the application.
Nonetheless, we try to make the focus of analysis as big as possible, so that the effect of other node activity in the averaged power consumption is negligible.

\subsection{Alya}\label{secAlya}

\paragraph{Introduction}
Alya~\cite{alya} is a high performance computational mechanics code to solve complex coupled multi-physics problems, discretized with Finite Element methods.
This software has been developed by the CASE Department of the Barcelona Supercomputing Center.
It is a modular scientific code written in Fortran and parallelized with MPI that is divided into multiple modules.
The particular software modules that are active during execution vary depending on the physics involved in the simulation.
In this work, we use the simulation of incompressible flows (\texttt{nastin}) and particle transport (\texttt{partis}).

\paragraph{Input}
The presented problem solves the simulation of airflow through the respiratory system extending up to approximately~$17$ bronchial generations and the transport of particles using a Lagrangian one-way coupling strategy.
The simulated steady airflow corresponds to a constant flow rate of~$30$~L/min passing through the inlet boundary.
The initial mesh consists of approximately~$45$M cells.

To enhance simulation accuracy and increase the problem size, the mesh can be refined.
Each refinement step subdivides the mesh by introducing intermediate nodes along the edges, resulting in an eightfold increase in the number of elements.
A single subdivision produces a mesh with roughly~$362$M elements, while two subdivisions yield a mesh with approximately~$2.90$B elements.
This mechanism is applied during the initialization phase of Alya and allows us to obtain a problem size that can scale to a higher number of nodes.

For the particle transport simulation,~$550$K particles, each with a diameter of~$120$~nm (representing the size of the SARS-CoV-2 virus), are injected at the start of the simulation (${t=0}$~s) and transported until they reach the walls, where they are considered deposited.
The number of particles remains constant regardless of mesh refinement.
In this work, we present results for both one and two levels of subdivision, demonstrating the performance of Alya at these two distinct scales.

\paragraph{Methodology}
We compiled two builds of Alya:
{\em i)} performance, to measure timing and energy metrics; and
{\em ii)} instrumented, to gather hardware counters and other runtime information for building the efficiency metrics model.

Regardless of the build type, the simulation runs for~100 timesteps and we repeat each run three times.
The focus of analysis encompasses all the timesteps.
All the results presented in this section show the measurements from the best run for a given node count.

\begin{figure}[!htbp]
  \centering
  \includegraphics[width=\columnwidth]{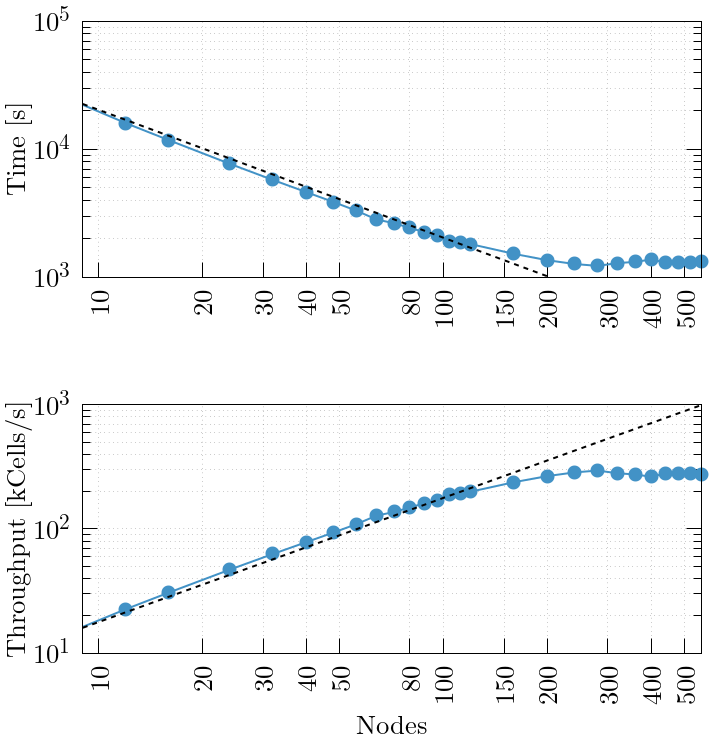}
  \caption{Elapsed time and throughput of Alya~($362$M cells) running in \mn5.}
  \label{figAlyaPerformance362}
\end{figure}

\paragraph{Scalability}
Figure~\ref{figAlyaPerformance362} shows the scalability of Alya with the~$362$M input when increasing the number of \mn5 nodes.
The plot at the top represents elapsed time and the plot at the bottom represents throughput measured in thousand cells per second (kCells/s) and computed as the number of simulated cells ($362$M) divided by the elapsed time divided by~$1000$.
Both plots include a dotted line which represents the ideal scalability with respect to the baseline run~(8 nodes).
The reader should note that axes are in logarithmic scale.

We observe a scalability that closely follows the ideal line, with some points slightly above, when running between~$10$ and~$100$ nodes.
This is because the work is distributed across more nodes, giving more memory resources and improving cache locality of each MPI rank.
The positive effects of this partitioning are present when running with higher node counts, but have diminishing returns as throughput deviates from the ideal starting at~$96$ nodes.
The throughput curve flattens completely from~$300$ nodes onward.

\begin{figure}[!htbp]
  \centering
  \includegraphics[width=\columnwidth]{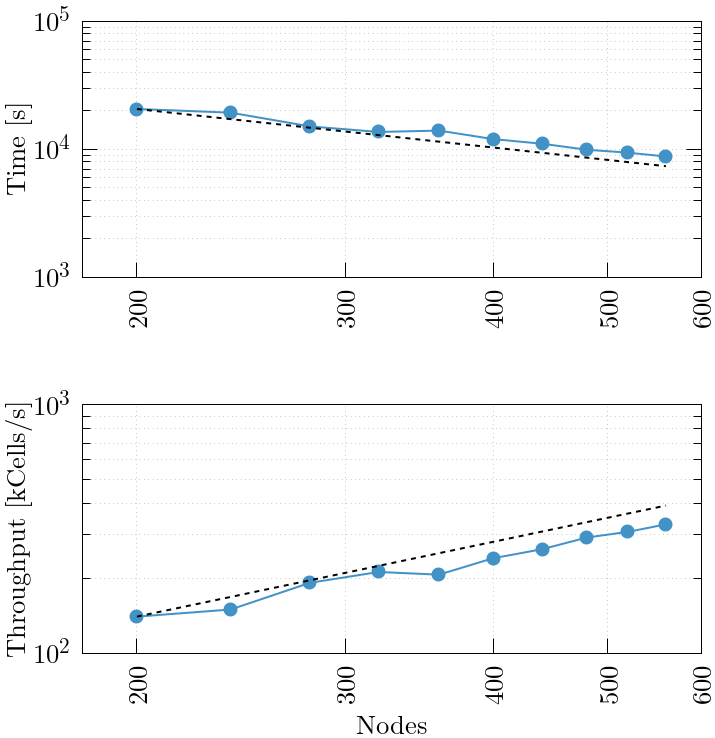}
  \caption{Elapsed time and throughput of Alya~($2.90$B cells) running in \mn5.}
  \label{figAlyaPerformance290}
\end{figure}

Figure~\ref{figAlyaPerformance290} shows the same scalability plots but running with the input with~$2.90$B cells.
In this case, the baseline run is with~200 nodes, since it is the minimum number of nodes required to run due to memory size constraints.

The cache locality effects also contribute to the scalability of the~$2.90$B input, although there are other limiting factors that we will now discuss.

\begin{figure*}[!htbp]
  \centering
  \includegraphics[width=\linewidth]{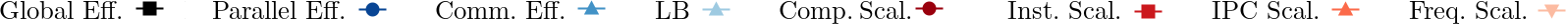}
  \includegraphics[width=\linewidth]{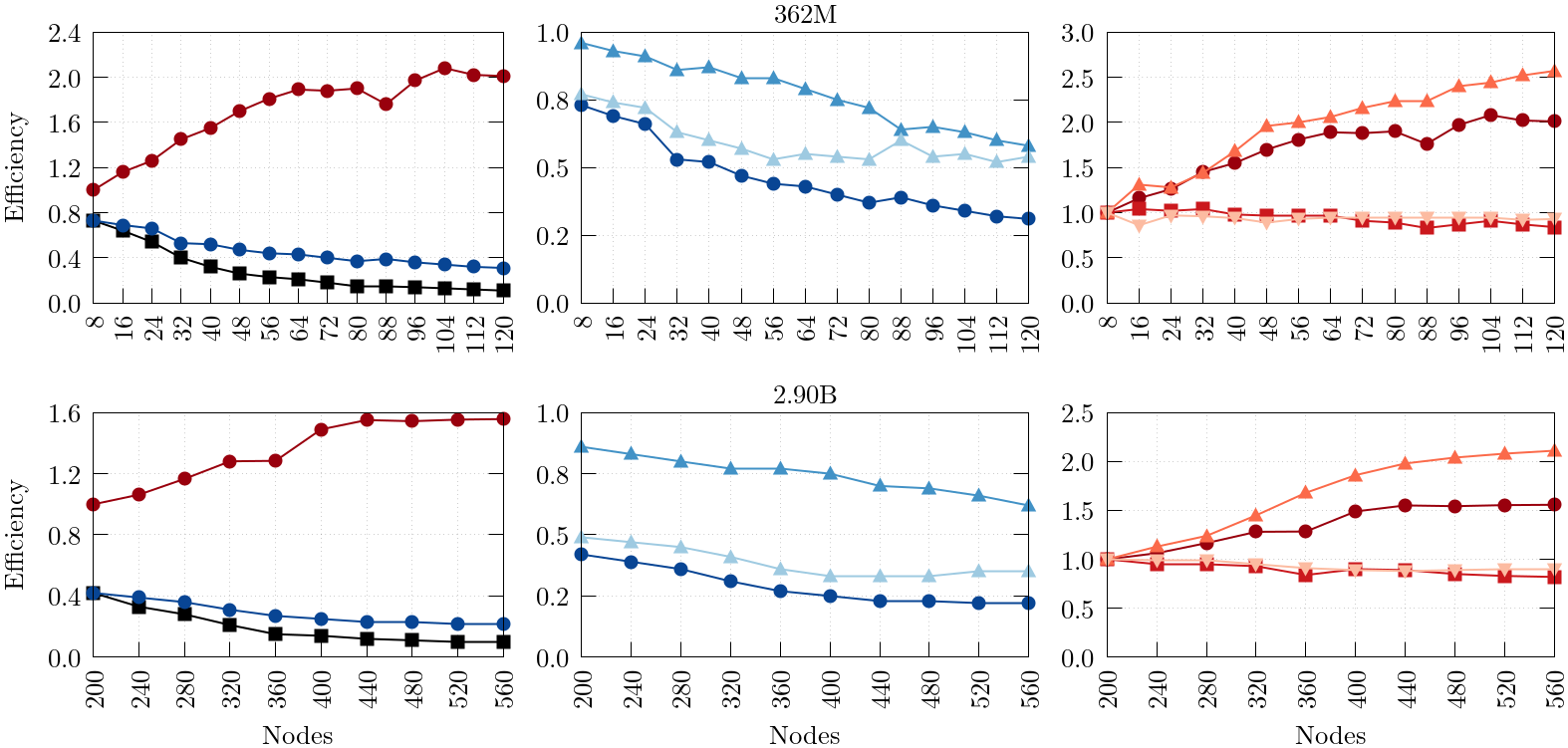}
  \caption{Efficiency metrics of Alya running in \mn5.}
  \label{figAlyaTalp}
\end{figure*}

\paragraph{Efficiency metrics}
Figure~\ref{figAlyaTalp} shows the efficiency metrics for both problem sizes.
The leftmost plot shows the metrics at the top level of the model;
the plot on the center shows the metrics underneath {\em Parallel Efficiency};
and the plot on the right shows the metrics underneath {\em Computational Efficiency}.

The {\em Parallel Efficiency} is always under~$0.75$ and decreases as the node count increases.
This is due to the decrease of both {\em Communication Efficiency} and {\em Load Balance}.
Of the two metrics, the {\em Communication Efficiency} is the one that drops de most for the~$362$M input;
starting at~$0.96$ with eight nodes and decreasing to~$0.58$ with~120 nodes.

The {\em Load Balance} is the main limiting factor since it is always the metric with the lowest value.
The cause of the imbalance is not discussed in this work and could be for a variety of reasons which include:
{\em i)} uneven workload distribution, in which some ranks have to do more work;
{\em ii)} imbalance of hardware resource usage, in which all ranks do similar amount of work, but some ranks perform it faster due to higher IPC.

We also observe a {\em Computational Scalability} which is always above~$1$.
This is mainly due to a previously discussed fact: higher node counts yield better memory locality, achieving better IPC.
The {\em Frequency Scalability} also stays high with values in the range of~$0.9$ to~$1.0$.
On the other hand, the {\em Instruction Scalability} drops under~$0.9$ starting at~$80$ nodes for the~$362$M input.

\begin{figure}[!htbp]
  \centering
  \includegraphics[width=\columnwidth]{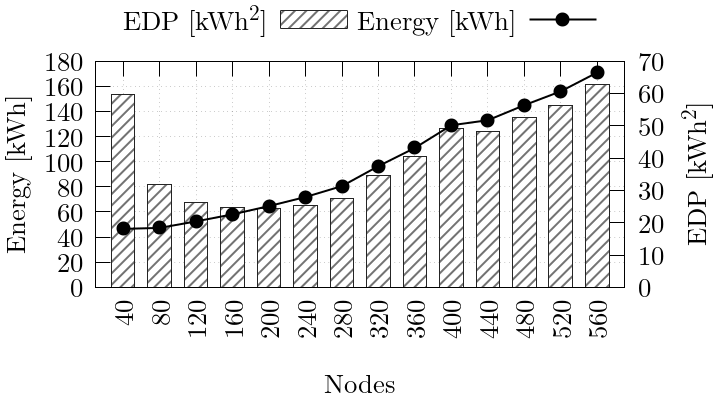}
  \caption{Total energy consumption and EDP of Alya~($362$M cells) running in \mn5.}
  \label{figAlyaEnergy362m}
\end{figure}

\begin{figure}[!htbp]
  \centering
  \includegraphics[width=\columnwidth]{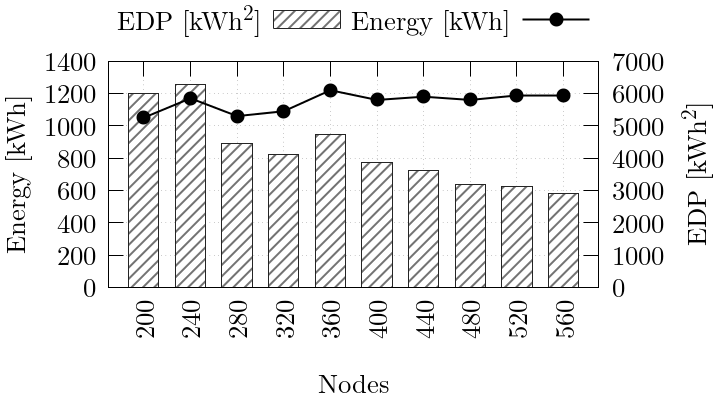}
  \caption{Total energy consumption and EDP of Alya~($2.90$B cells) running in \mn5.}
  \label{figAlyaEnergy290b}
\end{figure}

\paragraph{Energy consumption}
Figures~\ref{figAlyaEnergy362m} and~\ref{figAlyaEnergy290b} show the total energy consumption and energy-delay-product of Alya when running with each of the problem sizes.
The $x$-axis represents the number of nodes and some runs are omitted for clarity.

In the case of the input with~$362$M cells, we observe a linear increase in energy consumption as we increase the number of nodes.
Also, we observe that the EDP follows a U-shape.
By comparing these plots with the ones in Figures~\ref{figAlyaPerformance362} and~\ref{figAlyaPerformance290}, we make two observations.

First, the energy consumption stays flat when the application scales close to the ideal ($2.90$B cells in Figure~\ref{figAlyaEnergy290b}), but it increases close to linearly with the number of nodes when the application does not scale ($362$M cells with more than~100 nodes in Figure~\ref{figAlyaEnergy362m}).

Secondly, the EDP follows a U-shape:
When the application scales close to ideal, the EDP decreases as we increase nodes.
The absolute minimum of the EDP curve corresponds to the point in which the performance increase gained by adding more hardware resources does not compensate for the energy consumption cost.
This breaking point can be seen in Figure~\ref{figAlyaEnergy362m} and happens around~160 nodes, which is when the throughput curve in Figure~\ref{figAlyaPerformance362} deviates from the ideal and the {\em Global Efficiency} is below~$0.35$.

\subsection{OpenFOAM}\label{secOpenFOAM}

\paragraph{Introduction}



OpenFOAM~\cite{OpenFOAM} is a widely used free and open-source Computational
Fluid Dynamics (CFD) software that provides an extensive range of features for
solving complex fluid flow problems. The version used in this analysis,
OpenFOAM v2312, includes a new coherent file format for parallel I/O of
unstructured polyhedral meshes~\cite{OpenFOAM-coherent}, which optimizes data
access and reduces pre- and post-processing times. This approach improves
scalability and performance on large HPC systems, making OpenFOAM more
efficient for handling massive datasets and multi-node executions. This
development is part of the work driven by the exaFOAM project~\cite{exaFOAM},
aimed at enhancing OpenFOAM's scalability on modern HPC architectures, with
these improvements being validated using small-scale microbenchmarks, followed
by benchmarks, and large-scale HPC Grand Challenges, demonstrating its impact
across a range of scales and applications.

\paragraph{Input}

The OpenFOAM case selected is the exaFOAM Grand Challenge test case of the High
Lift Common Research Model (CRM-HL)~\cite{OpenFOAM-GC_Repo}, a full aircraft
configuration with deployed high-lift devices using wall-modelled LES (WMLES).
This case simulates the airflow around an aircraft in a landing configuration.

The geometry includes inboard and outboard leading-edge slats, single-slotted
flaps, and simplified flow-through engine nacelles, adding geometric complexity
and meshing challenges. The mesh used for this case consists of 1.79 billion
cells.

\paragraph{Methodology}

The simulation is executed for 200 timesteps, with the first timestep excluded
from the measurements as it involves initialization processes. Each run is
repeated five times to ensure reproducibility, and results from the best run on
a given node count are presented.

\begin{figure}[!htbp]
  \centering
  \includegraphics[width=\columnwidth]{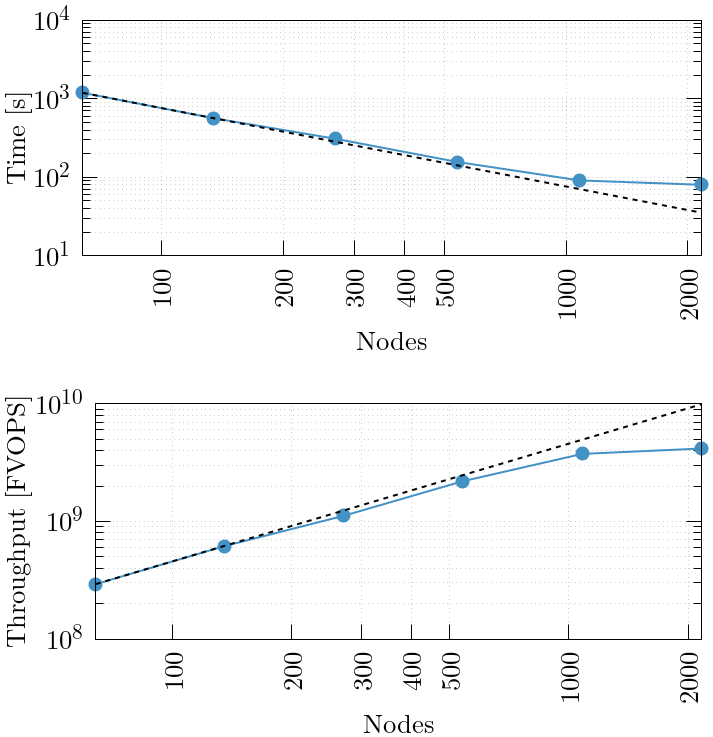}
  \caption{Elapsed time and throughput of OpenFOAM running in \mn5.}
  \label{figOpenFOAMPerformance}
\end{figure}

\paragraph{Scalability}

Figure~\ref{figOpenFOAMPerformance} shows the scalability of OpenFOAM as the
number of \mn5 nodes increases. The top plot represents the elapsed
time, while the bottom plot shows the throughput, measured in Finite Volumes solved Per Second (FVOPS)~\cite{FVOPS}. FVOPS is calculated based on
the number of cells in the grid processed during the time taken for a single
timestep (in seconds), offering a measure of the system's efficiency for each
individual computational step. Both plots include a dotted line representing
the ideal scalability with respect to the baseline run (64 nodes).

As the number of nodes increases, the execution time decreases, showing
significant performance improvements, especially between 64 and 540 nodes.
However, beyond 540 nodes, the reduction in execution time becomes less
pronounced. This trend is clearer in the throughput plot, where the measured
throughput in FVOPS closely follows the ideal scalability up to 540 nodes.
However, at 1080 nodes, the actual throughput begins to diverge from the ideal,
and at 2160 nodes the actual FVOPS is approximately half of the ideal value.

\paragraph{Efficiency metrics}

\begin{figure}[!htbp]
  \centering
  \includegraphics[width=\linewidth]{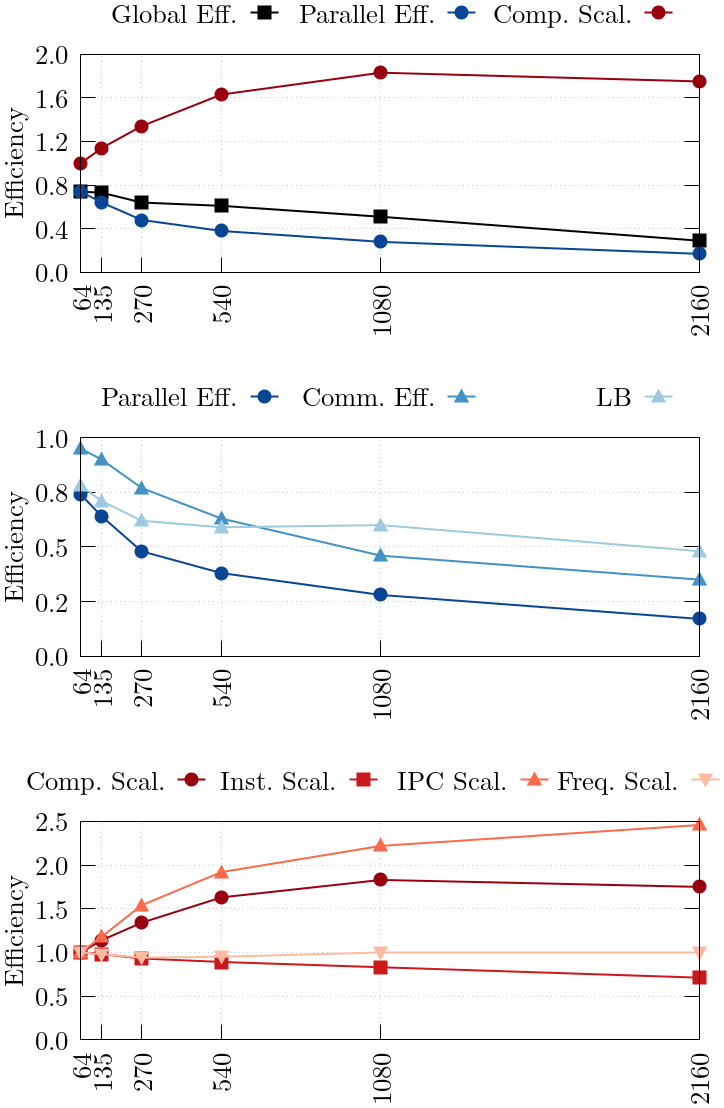}
  \caption{Efficiency metrics of OpenFOAM running in \mn5.}
  \label{figOpenFOAMTalp}
\end{figure}

Figure~\ref{figOpenFOAMTalp} presents the efficiency metrics for OpenFOAM for
the Grand Challenge case. The leftmost plot shows the top-level metrics, the
center plot focuses on the metrics under {\em Parallel Efficiency}, and the
rightmost plot presents the metrics under {\em Computation Efficiency}.

The {\em Parallel Efficiency} of OpenFOAM decreases significantly as the number
of nodes increases, starting at 0.74 with 64 nodes and dropping to 0.17 with
2160 nodes. This decline is primarily due to the reduction in {\em
Communication Efficiency}, which falls from 0.95 at 64 nodes to 0.35 at 2160
nodes. As more nodes are added, the ratio of inner cells to boundary faces
decreases, meaning that processors spend more time exchanging data rather than
performing computation. This increased communication overhead between processes
leads to a decrease in {\em Communication Efficiency}, causing an imbalance where the
time spent managing communications outweighs the time spent computing.

{\em Load Balance} experiences an initial drop as the number of nodes
increases, starting at 0.78 with 64 nodes and falling to 0.60 with 540 and 1080
nodes. It then drops further to 0.48 with 2160 nodes. This decline in {\em Load
Balance} can be attributed to the uneven distribution of computational work
among the processing nodes. Adding more nodes, compute-intensive regions
of the simulation become sparsely distributed and separated from regions that
require less computational effort and converge more quickly, resulting in some
processors handling significantly more workload than others.

On the other hand, {\em Computation Scalability} increases with the number of
nodes, peaking at 1.83 with 1080 nodes. This high efficiency is attributed to
the good {\em IPC Scalability} of the case. IPC increases with more nodes,
indicating that with fewer cells per core, memory bandwidth becomes less of a
bottleneck, resulting in better memory locality. This observation aligns with
findings from prior works~\cite{OpenFOAM-memory-bandwidth}, which identified
that the region under study in OpenFOAM is memory bandwidth bound.

The superlinear {\em Computation Scalability} compensates the low {\em Communication Efficiency} up to~1000 nodes.
Beyond this point, the limited {\em Instruction Scalability} outweighs the increase in~IPC, making it so the {\em Global Efficiency} drops.

\begin{figure}[!htbp]
  \centering
  \includegraphics[width=\columnwidth]{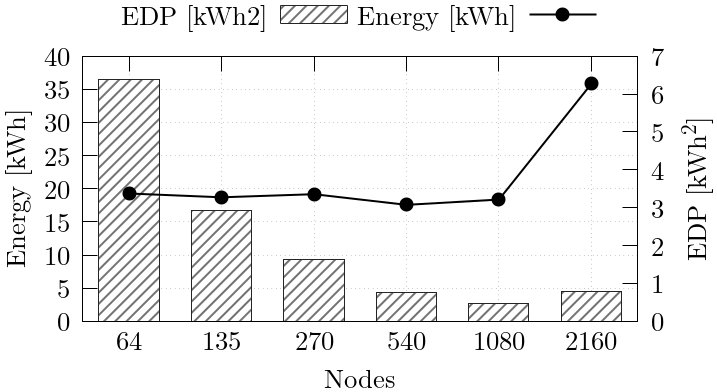}
  \caption{Total energy consumption and EDP of OpenFOAM running in \mn5.}
  \label{figOpenFOAMEnergy}
\end{figure}

\paragraph{Energy consumption}
Figure~\ref{figOpenFOAMEnergy} presents the total energy consumption and
energy-delay-product of OpenFOAM for all executions.
We observe that, up to~1080 nodes, energy usage remains stable, fluctuating slightly but staying
within a narrow range. At 2160 nodes, there is a significant increase
in energy consumption.

The EDP curve decreases as the number of nodes increases, reaching its lowest
point at 1080 nodes. Beyond this point, the EDP rises again at 2160 nodes.

OpenFOAM shows a constant energy consumption and a decreasing EDP up to~1080
nodes, where it scales efficiently. However, at~2160 nodes, both energy
consumption and EDP increase, matching the flattening of the curve
observed in Figure~\ref{figOpenFOAMPerformance}.

\subsection{IFS}

\paragraph{Introduction}

The Integrated Forecasting System (IFS)~\cite{wedi2020baseline} atmospheric
model is an operational global meteorological forecasting model developed by the
European Centre for Medium-Range Weather Forecasts (ECMWF). The dynamical core
of IFS is hydrostatic, two-time-level, semi-implicit, semi-Lagrangian and
applies spectral transformations between grid-point and spectral space. This
model is used for daily weather predictions by ECMWF, and its results serve as
the foundation for forecasts issued by major meteorological agencies across
Europe.

IFS can also be employed for climate studies when combined with simulations of
other components, such as the ocean. These studies require high-resolution
global simulations, which involve significant computational costs and would be
impossible without supercomputers.

The model is adapted for HPC environments, written in Fortran, and optimized for
parallel processing through a hybrid combination of MPI and OpenMP. The version
used in this work (CY48R1) is also currently used in a production environment.

\paragraph{Input}

The configuration set for the tests is used for climate prediction, starting
from the year~2020 onward. This configuration is based on real observational
forcing data, including variables such as wind and CO2 emissions from~1950, to
achieve a stable and realistic setup for predictions from~2020 onward.

For the results presented in this paper, a~10-day simulation was performed with
a horizontal resolution of 4~km and~137 vertical levels. The NPROMA (block
processing size) used for the evaluation runs was set to~16. 

\begin{figure}[!htbp]
  \centering
  \includegraphics[width=\columnwidth]{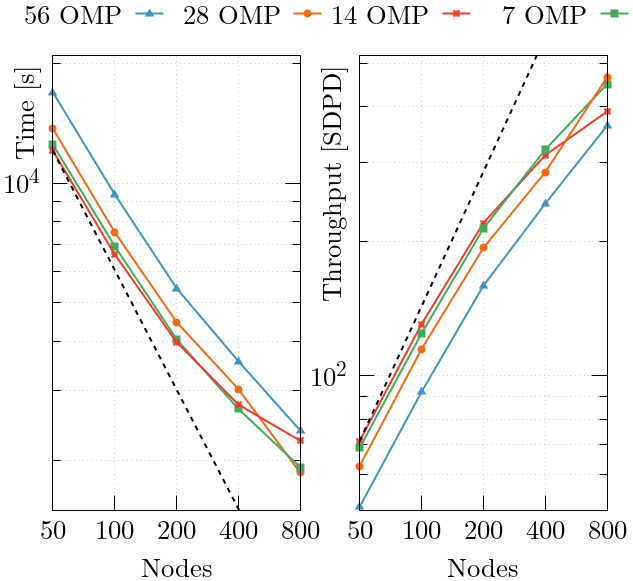}
  \caption{Elapsed time and throughput of IFS running in \mn5.}
  \label{figIFSPerformance}
\end{figure}

\paragraph{Methodology}
We compile IFS with the ifs bundle from ECMWF. 
The model runs for~10 simulated days and we repeat each run
three times.
The time ($T_{IFS}$, in seconds) of each run is measured by aggregating the
execution time of each timestep as reported by IFS (removing the first 3
timesteps, which contain initialization).
To compute the throughput (SDPD, Simulated Days Per Day), we divide the
simulated time in days ($9.992188$ days, after removing 3 timesteps) by the
execution time in days.
%
\paragraph{Scalability}
Figure~\ref{figIFSPerformance} shows the scalability of IFS when increasing the
number of \mn5 nodes in the $x$-axis.
The different series plotted represent different combinations of MPI processes and OpenMP threads.
All the runs of Figure~\ref{figIFSPerformance} use the whole node, therefore, the
amount of MPI processes per node is 112 divided by the amount of OpenMP threads.
The plot at the top shows elapsed time (in logarithmic scale for
both axes) and the plot at the bottom displays throughput.
Both plots include a dotted line which illustrates the ideal scalability with
respect to the best MPI-OpenMP configuration with 50 nodes.

In this Figure we can observe that when running in 50, 100, 200 and 400 nodes
the best combination of MPI-OpenMP is to use~7 or~14 OpenMP threads for each MPI process.
However, when running in 800 nodes, using 28 OpenMP threads per MPI ranks is the best option.
We can conclude that having less OpenMP threads is better for lower amounts of nodes. However,
we see that for the higher amounts of nodes trading processes for threads
increases the throughput. We observe that the MPI scalability stalls at around
10k MPI ranks, and only then its useful to increase the amount of OpenMP
threads.

\begin{figure}[!htbp]
    \centering
    \includegraphics[width=\columnwidth]{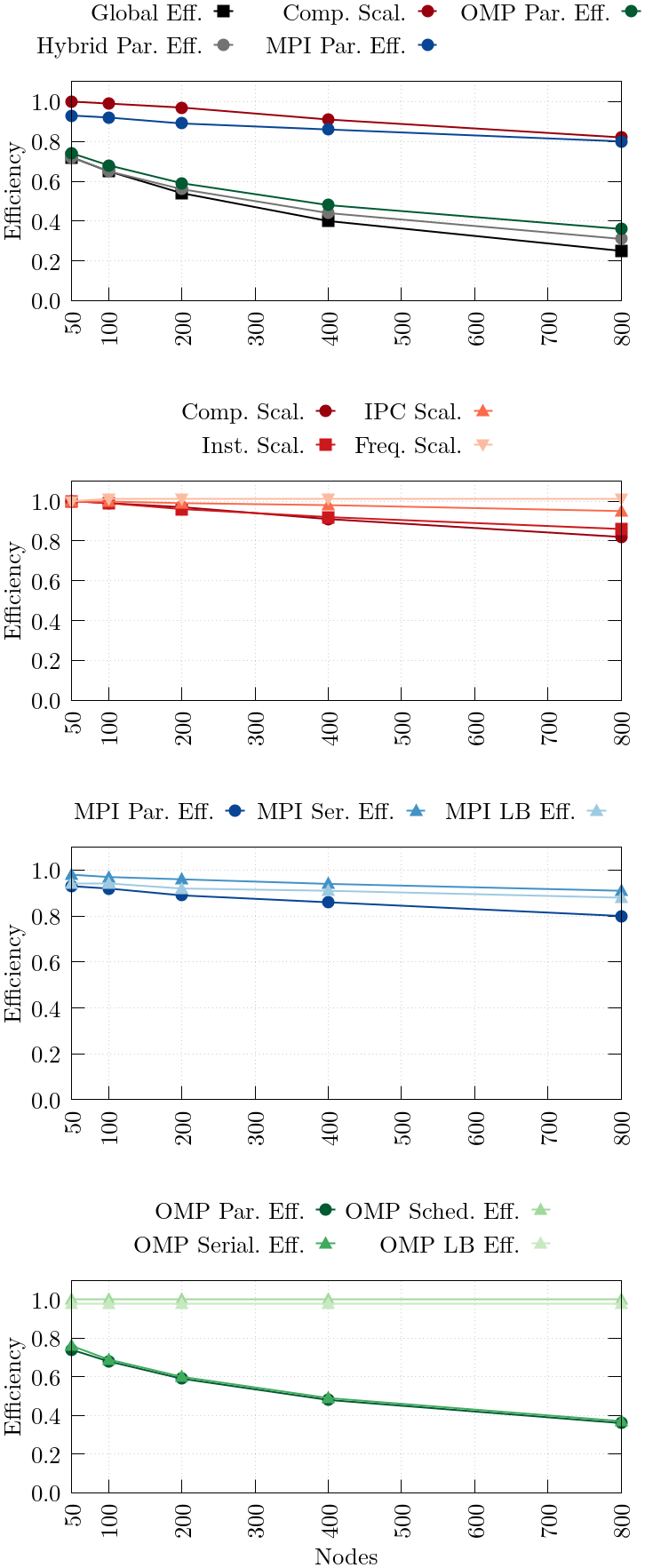}
    \caption{Efficiency metrics of IFS running in \mn5.}
    \label{figIFSTalp}
\end{figure}

\paragraph{Efficiency metrics}
We have measured the IFS efficiency metrics only for one MPI-OpenMP combination
(8 MPI x 14 OpenMP)\footnote{we have picked this combination, because we thought
it would be the combination that would give more insight for both MPI and OpenMP
efficiencies}. Figure~\ref{figIFSTalp} shows the efficiency metrics, and for
this application we include the OpenMP efficiencies.
From top to bottom, Figure~\ref{figIFSTalp} is read as follows:
{em i)} parent metrics of all the other plots, with the addition of the {\em Hybrid Parallel Efficiency} and {\em Global Efficiency};
{\em ii)} breakdown of the {\em Computation Scalability};
{\em iii)} breakdown of the {\em MPI Parallel Efficency}; and
{\em iv)} breakdown of the {\em OpenMP Parallel Efficiency}.

Looking at the top chart we can see that IFS have a good Computation Scalability and MPI efficiency, being both metrics above~$0.80$ for all the cases. On the other hand, the global efficiency and the hybrid parallel efficiency are very low, with values always below~$0.80$ and reaching~$0.30$ when running in 800 nodes. These low values are due to the lo OpenMP parallel efficiency which is the metric dragging the other ones.

Looking at the Computational Scalability metrics (red lines) we can see that the IPC and frequency scalability show very good scaling values always above~$0.90$. Only the Instruction Scalability is the worse one but still showing very good values always above~$0.80$, this metric measures the amount of replicated code and is not surprising to see it decrease with this large number of core counts, usually meaning that the amount of work assigned to each MPI rank is getting smaller and the computational of boundaries or replicated code is becoming relevant.
For the MPI metrics (blue lines) we can observe that {\em MPI Serialization Efficiency} and {\em MPI Load Balance Efficiency} both have nearly the same impact on their parent metric.
However, as mentioned in Appendix~\ref{secMetrics}, the inefficiencies introduced by MPI fall into OpenMP metrics in the Hybrid model (\ie {\em OpenMP Serial Efficiency}). Knowing the nature of the earth models load imbalance (the load distribution imbalance is not constant across timesteps due to changes in solar
radiation and atmospheric phenomena such as precipitation)~\cite{balancing_ec_earth3}, these kind of irregular imbalances are accounted under {\em MPI Serialization Efficiency}. So we can conclude that the main issue when scaling MPI is the load imbalance, either from a global point of view or because of micro-imbalances.
When we analyze the OpenMP metrics (green lines) we can see that the OpenMP scheduling efficiency and OpenMP Load Balance are almost perfect, and the low OpenMP efficiency is due to OpenMP Serialization. OpenMP serialization is a metric that indicates the amount of code parallelized with OpenMP, in this case we can assume that the OpenMP parallelization of the code is not enough, this is also backed up by the scaling data that we have seen before, were it was better to use less OpenMP threads and more MPI ranks.

Moreover, another factor that impacts the serialization metric is MPI blocking calls that are not overlapped with parallel OpenMP work, this is also considered OpenMP Serialization, for this reason we see a decrease of the OpenMP Serialization metric when increasing the number of nodes, even though the number of OpenMP threads per process is constant.

Combining both the results from this metrics and the scalability we can tell
that, overall, IFS performance is driven by time spent in MPI calls and low
OpenMP parallelization.

\paragraph{Energy consumption}

\begin{figure}[!htbp]
  \centering
  \includegraphics[width=\columnwidth]{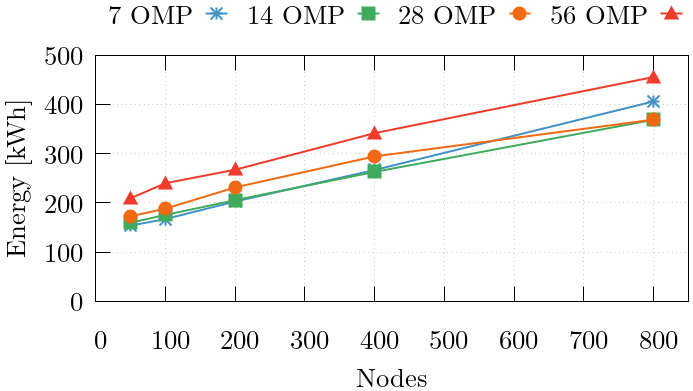}
  \caption{Total energy consumption of IFS running in \mn5.}
  \label{figIFSEnergy}
\end{figure}

Figure~\ref{figIFSEnergy} show the total energy consumption of IFS for the
different MPI-OpenMP combinations. The energy consumption grows linearly
correlated to the scalability efficiency.

\begin{figure}[!htbp]
  \centering
  \includegraphics[width=\columnwidth]{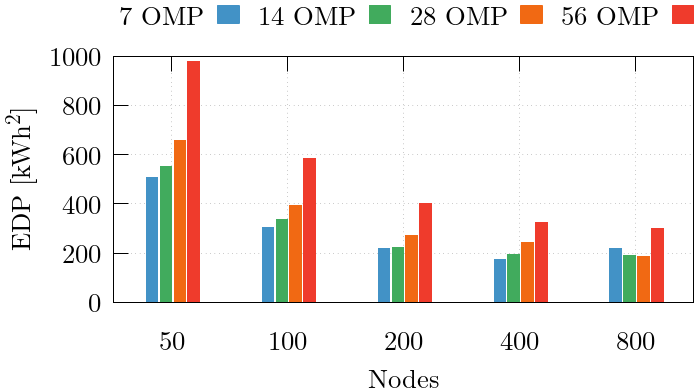}
  \caption{EDP of IFS running in \mn5.}
  \label{figIFSEDP}
\end{figure}

Figure~\ref{figIFSEDP} shows the energy-delay-product of IFS for the different
MPI-OpenMP combinations. The 7 OpenMP threads executions draw a U-shape, with
its minimum at 400 nodes. This plot also shows that when looking for the best
energy-throughput trade-off it makes more sense to run with lower number of
OpenMP threads, for example at 400 nodes 7 OpenMP threads shows better EDP
compared to 14 OpenMP threads, even through the 14 OpenMP threads run was
faster. Its with 800 nodes that it starts to make sense energetically to use
more OpenMP threads.

\section{Memory Technology Evaluation}\label{secHbm}

In this section we evaluate the HBM nodes of the general purpose partition in \mn5.
Although they share the same core micro-architecture and core count with the DDR nodes,
these nodes include~$8$ dies of~$16$~GB of HBM2 memory, adding up to~$64$~GB.
HBM nodes have a peak memory bandwidth of~$3.28$~TB/s, which is~$5.33\times$ higher than the peak of a DDR node.

In addition to the sustained bandwidth, we try to measure the power efficiency of the HBM nodes.
Unfortunately, the power monitoring infrastructure does not provide a mechanism to measure the consumption of HBM in isolation.
Due to this limitation, we only measure the power consumption of the whole node.

\subsection{Low-level Benchmarks}

\begin{figure}[!htbp]
  \centering
  \includegraphics[width=\columnwidth]{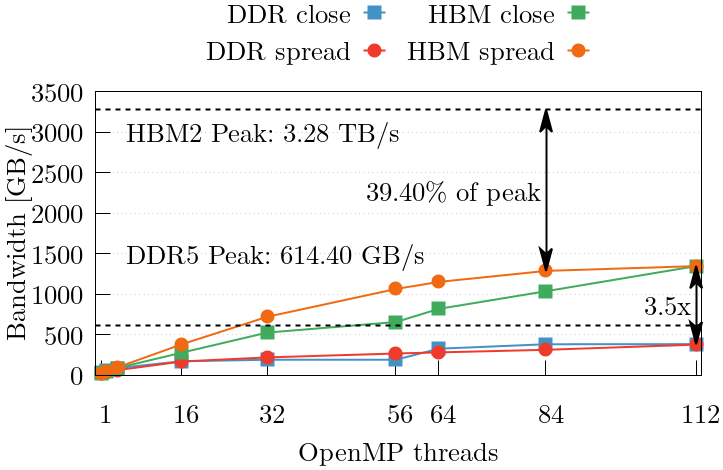}
  \caption{Aggregated copy bandwidth in DDR and HBM nodes.}
  \label{figHbmMemKernel}
\end{figure}

\paragraph{\mem}
We repeat the same experiment as in Section~\ref{secBandwidth} in which we run the \mem with a code in which each OpenMP thread copies and array of~$64$~MiB from one memory location to another.
We also run with two binding policies, \texttt{close} and \texttt{spread}.
Figure~\ref{figHbmMemKernel} shows the sustained bandwidth measured with \mem comparing the DDR and HBM nodes, and the two binding policies.

Our measurements show that the \texttt{close} policy in the HBM nodes behaves similarly compared to the DDR nodes.
Both curves (blue and green) have two steps which correspond to the bandwdith saturation of the first and second socket, respectively.
On the other hand, the \texttt{spread} policy is somewhat different in the HBM nodes.
It is similar to the one in DDR nodes in that it has a single smooth step.
However, the sustained bandwidth is always higher than the \texttt{close} policy.
This was not the case when running in the DDR nodes between~$64$ and~$112$ OpenMP threads.

Looking at the maximum aggregated bandwidth, we observe that the HBM nodes achieve~$1.35$~TB/s, which is~$3.5\times$ more memory bandwidth than in the DDR nodes.
Nonetheless, this bandwidth is very far from the theoretical peak, achieving only~$39.40\%$.
Our experiment shows that the use of HBM memory is beneficial for a memory bound code in raw bandwidth numbers, the Sapphire Rapids CPU cannot fully leverage all the bandwidth that the HBM can provide.

\begin{figure}[!htbp]
  \centering
  \includegraphics[width=\columnwidth]{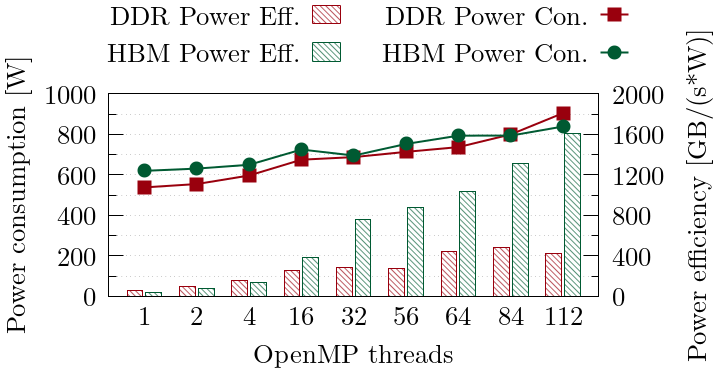}
  \caption{Node power consumption in DDR and HBM nodes.}
  \label{figHbmMemPower}
\end{figure}

\paragraph{Power Efficiency}
Figure~\ref{figHbmMemPower} shows the node power consumption and efficiency of DDR and HBM nodes when running the \mem with a full node ($112$~OpenMP threads).
The $x$-axis represents the number of threads,
the left $y$-axis represents the node power consumption, and
the right $y$-axis represents the power efficiency.
The reader should note that the $x$-axis is neither linear nor logarithmic scale.

Our measurements show that the HBM nodes consume around the same power compared to the DDR nodes.
Runs with small core counts (\aka up to~$32$ threads) consume more in the HBM nodes, probably because the idle power consumption is higher.
Considering a similar power consumption but a~$3.5\times$ increase in memory bandwidth, the HBM nodes achieve a much higher power efficiency than the DDR counterparts.

\begin{figure}[!htbp]
  \centering
  \includegraphics[width=\columnwidth]{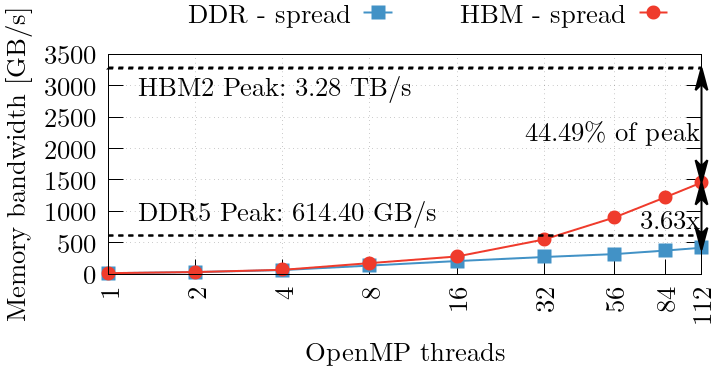}
  \caption{Sustained memory bandwidth running STREAM in \mn5 nodes with and without HBM. Dotted lines represent the theoretical peak.}
  \label{figStreamHbm}
\end{figure}

\paragraph{STREAM}
We repeat the executions of STREAM in the HBM nodes to compare their performance with the DDR nodes (see Section~\ref{secBandwidth}).
Figure~\ref{figStreamHbm} compares the memory bandwidth achieved by the Triad kernel when using the DDR and HBM nodes.

In line with our results with the \mem, the sustained bandwidth when using HBM memory is much higher compared to DDR.
At best, our measurements show an $3.63\times$ improvement of HBM over DDR5.
However, it is also important to note that the bandwidth achieved in the HBM configuration is very far from its theoretical peak.
It only reaches up to $44.49\%$ of the peak.
This is due to the limitation on the CPU to leverage the huge bandwidth that the HBM memory offers and has also been discussed by McCalpin~\cite{10.1007/978-3-031-40843-4_30}.

\begin{tabox}
  \begin{itemize}[leftmargin=*]
    \item HBM nodes provide up to~$3.63\times$ more memory bandwidth than DDR nodes
    \item HBM nodes have a similar power consumption than DDR nodes, which means that they provied higher power efficiency
    \item The Sapphire Rapids can only leverage up to~$44.49\%$ of the peak HBM2 memory bandwidth
  \end{itemize}
\end{tabox}

\subsection{Alya}

In this section, we compare the execution of Alya in one node of \mn5 with and without HBM memory.
Due to memory constraints, we run a different input than presented in Section~\ref{secAlya}.
For the DDR and HBM comparison, the input simulates a synthetic sphere mesh of~$16$M cells.
We run the simulation for~750 timesteps and measure the total execution time as well as the energy and EDP using the same methodology explained in Section~\ref{secMethodology}.

\begin{figure}[!htbp]
  \centering
  \includegraphics[width=\columnwidth]{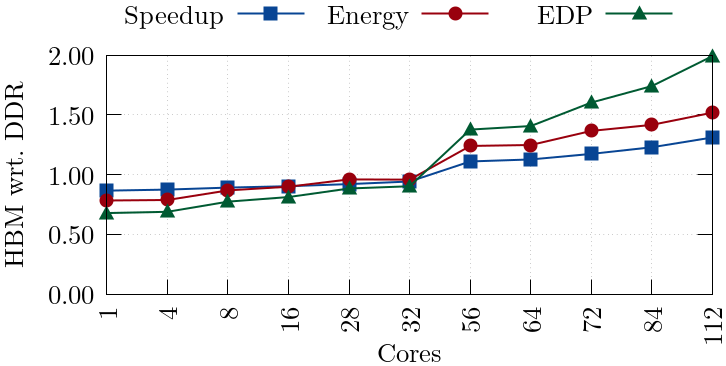}
  \caption{Speedup, energy and EDP of HBM nodes with respect to DDR nodes in \mn5 when running Alya (\texttt{close}).}
  \label{figAlyaHbm}
\end{figure}

Figure~\ref{figAlyaHbm} shows the measurements of the HBM node normalized to the ones in DDR.
Values under one mean that the respective metric is lower in the DDR nodes compared to HBM.
For speedup, higher is better;
while for energy and EDP, lower is better.
Moreover, each point in the curve of EDP is mathematically equivalent to the product of the points in the speedup and energy curves.

The results show that for up to~32 cores, the HBM node runs slightly slower, with speedup values below one.
Beyond~32 cores, the speedup increases, peaking at $1.31\times$ with a full node.

At core counts lower than~32, the HBM node consumes less energy compared to the DDR node.
However, the difference between the two node types decreases as the core count increases.
With~112 cores, energy consumption in the HBM configuration is~$1.52\times$ more than
the DDR configuration.

The EDP follows a similar pattern: it is lowest at lower core counts in the HBM node, indicating better energy delay performance.
However, as the number of cores scales, the EDP increases significantly, reaching a peak of~$1.99\times$ at 112 cores, double the one of DDR nodes.

\subsection{OpenFOAM}

This section compares the performance of OpenFOAM on a single node of \mn5 when
using DDR and HBM memory.

For this experiment, we use a microbenchmark representative of the use case
used in Section~\ref{secOpenFOAM}, referred as MB9 in the work by Galeazzo
et.al~\cite{CFDcomparisons}. This case, referred to as the High-Lift
Configuration, is a compressible Large Eddy Simulation (LES) benchmark
simulating the aerodynamic behavior of a wing with complex flow patterns. The
mesh for this case consists of 19.5 million cells.

The simulation is executed for different number of cores inside a node, running
20 timesteps for each configuration and discarding the first timestep to
exclude initialization processes.

\begin{figure}[!htbp]
  \centering
  \includegraphics[width=\columnwidth]{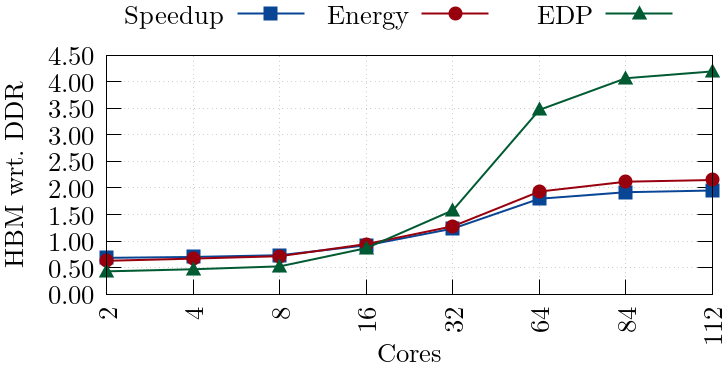}
  \caption{Speedup, energy and EDP of HBM nodes with respect to DDR nodes in
    \mn5 when running OpenFOAM (\texttt{spread}).}
  \label{figOpenFOAMHbm}
\end{figure}

We report the speedup in execution time, energy consumption, and EDP for the
HBM node, normalized to the DDR configuration. Figure~\ref{figOpenFOAMHbm}
shows these metrics, where values below one indicate that the respective metric
is lower on the DDR nodes compared to the HBM nodes.

The results show that for up to~16 cores, the HBM node runs slightly slower, with speedup values below one.
Beyond~16 cores, the speedup increases significantly, peaking at $1.95\times$ at 112 cores, nearly doubling the performance.

At lower core counts, the HBM node consumes more energy.
However, the difference between the two node types decreases as the core count increases.
With~112 cores, energy consumption in the HBM configuration is approximately~$2.15\times$ more than the DDR configuration.

The EDP follows a similar pattern: it is lowest at lower core counts in the HBM node, indicating better energy delay performance.
However, as the number of cores scales, the EDP increases significantly, reaching a peak of~$4.19\times$ at~112 cores.

These results indicate that, although HBM gives better speedup for OpenFOAM at higher levels of parallelism, it also comes with increased energy consumption and EDP than DDR nodes.

\begin{tabox}
  \begin{itemize}[leftmargin=*]
  \item HBM nodes achieve~$1.31\times$ and~$1.95\times$ speedup for Alya and OpenFOAM, respectively, with respect to the same core counts in DDR nodes
  \item The improved speedup in HBM nodes does not compensate for the increase in energy consumption, yielding an increase of EDP of up to~$4.19\times$ compared to DDR
  \end{itemize}
\end{tabox}

\section{Conclusions}\label{secConclusions}
This document provides a detailed evaluation of \mn5, a pre-exascale supercomputer hosted at the Barcelona Supercomputing Center. The analysis encompasses the system’s architectural features, benchmarking at multiple levels, and the performance of key scientific applications. The document aims to complement traditional technical documentation by offering insights into \mn5’s behavior and capabilities, thus assisting users in optimizing their workflows.

The micro-benchmarking results reveal that \mn5 achieves performance near theoretical peaks for many low-level metrics. Floating-point performance tests confirm that Intel Sapphire Rapids CPUs in the General Purpose Partition (GPP) sustain at least 99.7\% of their maximum performance. Memory bandwidth evaluations show that DDR5 configurations achieve up to 62.87\% of their theoretical peak, a result consistent with similar architectures. The interconnect tests demonstrate a bandwidth of~$25$~GB/s for packages greater than 64~KiB and latency of~$1.5$~us.

Application benchmarks highlight the practical capabilities and challenges of using \mn5 for scientific workloads. Alya scales efficiently up to moderate node counts (around 100), with diminishing returns observed as communication overheads and load imbalance emerge at higher node counts. OpenFOAM demonstrates strong scalability up to approximately 1000 nodes but experiences performance degradation due to increased communication overheads and reduced computational efficiency beyond this point. IFS exhibits notable inefficiencies at large scales (more than~200 nodes), particularly due to load imbalance and serialization in MPI and OpenMP hybrid parallel configurations. These findings emphasize the importance of optimizing both code and workload distribution to fully leverage \mn5’s resources.

Energy efficiency, a key focus of \mn5’s design, is supported by the EAR (Energy Aware Runtime) framework. Direct liquid cooling effectively manages thermal loads, ensuring consistent performance and reducing energy consumption. Benchmarking results show that energy consumption scales nearly linearly with increased node usage for most applications, while the Energy-Delay Product (EDP) indicates that optimal configurations balance computational efficiency with energy use. Notably, for applications like Alya, there is a clear U-shaped EDP curve, with the lowest values observed at the scalability sweet spot.

The comparison between DDR5 and HBM configurations reveals significant differences in performance and energy consumption. HBM nodes achieve speedups of up to $3.6\times$ for memory-bound workloads, particularly in applications requiring high memory throughput. However, these gains come with increased energy costs at higher core counts, where HBM configurations consume approximately $2.1\times$ the energy of their DDR5 counterparts. This trade-off highlights the importance of selecting memory configurations based on the workload’s computational characteristics. For example, HBM is advantageous for highly memory-intensive applications, while DDR5 remains a more energy-efficient choice for less memory-demanding scenarios.

Overall, the evaluation underscores \mn5’s strengths and provides a comprehensive resource for understanding its performance characteristics. While the system achieves impressive results across various benchmarks, further work is needed to optimize application scalability and energy efficiency at larger scales. The findings herein will guide users in tailoring their workloads to maximize scientific output and make effective use of \mn5’s computational power.

\section*{Acknowledgments}

Supported by the pre-doctoral program AGAUR-FI ajuts (2024 FI-200424) Joan Oró offered by Secretaria d'Universitats i Recerca del Departament de Recerca i Universitats de la Generalitat de Catalunya.

The acquisition and operation of the EuroHPC supercomputer is funded jointly by the EuroHPC Joint Undertaking, through the European Union's Connecting Europe Facility and the Horizon 2020 research and innovation programme, as well as the Participating States Spain, Portugal, and T\"urkiye.


\appendix
%

\section{Efficiency Metrics}\label{secMetrics}

\subsection*{Base model}
\begin{figure*}[!htbp]
  \centering
  \includegraphics[width=\textwidth]{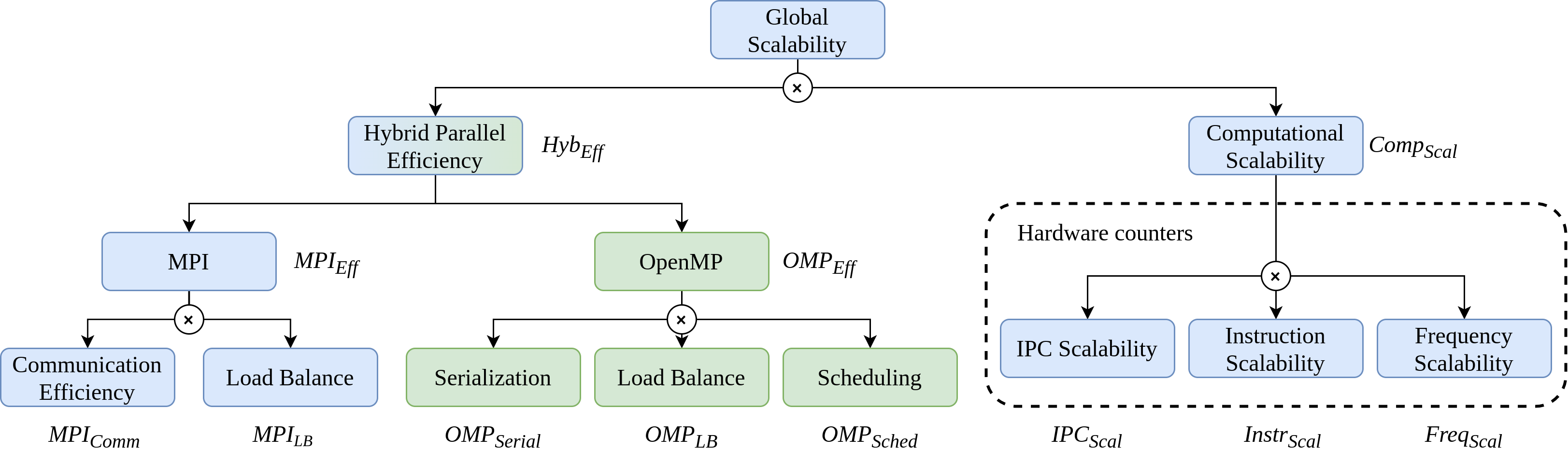}
  \caption{Efficiency metrics model.}
  \label{figEfficiencyMetrics}
\end{figure*}

This performance efficiency model is structured as a tree of metrics which is shown in Figure~\ref{figEfficiencyMetrics}.
All metrics with the exception of the ones labeled {\em Scalability} take values between zero and one.
The {\em Scalability} metrics are computed with respect to a reference run (\eg single-core execution) and can take values above one.

Metrics shown in blue boxes have been previously defined in~\cite{pop_metrics}.
Metrics shown in green boxes are part of a new hybrid efficiency model that is currently in development and we formally formulate them later in this section.

In the base model, all metrics can be computed by the product of their children metrics.
This is noted with~$\times$ in Figure~\ref{figEfficiencyMetrics}.
The hybrid model renames the {\em Parallel Efficiency} metric to {{\em Hybrid Parallel Efficiency}}, which accounts for both MPI and OpenMP, and defines an efficiency metric for each one of the programming models ($MPI_{Eff}$ and $OMP_{Eff}$).
This is the only instance of the hybrid model in which children metrics cannot be multiplied to compute the parent metric.

Lastly, some metrics require accessing hardware counters (cycles and instructions) to be computed.
{\em Computational Scalability}
represents the change of useful computation across different executions.
Ideally, the aggregated time of useful computation across all cores should remain constant for strong scaling applications.
The children metrics {\em Instruction, IPC, and Frequency Scalability} represent the change of their respective elements across different executions.
These four metrics are the only ones that require reading hardware counters,
and are highlighted with a dotted-line box in Figure~\ref{figEfficiencyMetrics}.
%

\subsection*{Extended hybrid model}
This extended model includes metrics for the OpenMP parallel programming model.
This work is the first one to formally formulate such metrics.
For simplicity, we assume that all MPI processes run with the same number of OpenMP threads,
and that all OpenMP threads execute the same parallel regions.
This assumtion holds true for IFS, which is the only hybrid application under study in this work.

Let $N$ be the number of MPI processes;
$M$ be the number of OpenMP threads per process;
and $P$ the number of parallel regions of each thread.

\paragraph{Classification of execution time}
Let $T_{i,j}$ be the total time spent in the $j$-th OpenMP thread of the $i$-th MPI process.
We classify time spent by threads in three categories:
{\em i)} $U$, doing useful computation;
{\em ii)} $C$, MPI communication; and
{\em iii)} $I$, idle OpenMP threads\footnote{We also consider time spent within the OpenMP runtime as "idle".}.
Let $T^{k}_{i,j}, \forall k \in \{U, C, I \}$ be the time spent in the $k$-th category for the given thread and process.
Therefore,
$$T_{i,j} = T^{U}_{i,j} + T^{C}_{i,j} + T^{I}_{i,j}$$

Let $T^k_{i}$ be the average thread time per process such that:
$$
T^{k}_{i} = \frac{\sum_{j}^{M} T_{i,j}^{k}}{M},~
\forall k \in \{U, C, I \}
$$

We further subdivide the time spent by idle threads ($T^I$) into three subcategories:
{\em i)} $I_{serial}$, outside parallel regions;
{\em ii)} $I_{lb}$, waiting for the slowest thread of each parallel region; and
{\em iii)} $I_{sch}$, all other time spent idling inside parallel regions.

Let $T^I_{i,j,p}$ be the idle time inside the $p$-th OpenMP parallel region of the $j$-th thread of the $i$-th process.

Let $T^{I_{serial}}_{i}$ be the average thread time spent by process $i$ outside OpenMP,
$$ T^{I_{serial}}_{i} = \frac{\sum_{j \in [1,M]} (T^{I}_{i,j} - \sum_{p \in [1,P]} T^I_{i,j,p})}{N} $$

Let $T^{I_{lb}}_{i}$ be the average thread time spent by process $i$ waiting for thread synchronization due to load imbalance,
$$ T^{I_{lb}}_{i} = \sum_{p \in [1,P]}( T^I_{i,p} - \min_{j \in [1,M]} T^I_{i,j,p} ) $$

Let $T^{I_{sch}}_{i}$ be the average thread time spent by process $i$ lost due to scheduling,
$$ T^{I_{sch}}_{i} = \sum_{p \in [1,P]} \min_{j \in [1,M]} T^I_{i,j,p} $$

Let $T^k$ be the average process execution time such that:
$$
T^{k} = \frac{\sum_{i}^{N} T_{i}^{k}}{N},~
\forall k \in \{U, C, I_{serial}, I_{lb}, I_{sch} \}
$$

\paragraph{OpenMP metrics}
Let $OMP_{Serial}$ ({\em OpenMP Serial Efficiency}) represent the time lost because OpenMP was not running a parallel region.
$$ OMP_{Serial} = (T - T^{I_{serial}}) / T $$
This metric is used to account for regions not parallelized with OpenMP.

Let $OMP_{LB}$ ({\em OpenMP Load Balance Efficiency}) represent the time lost in idle OpenMP threads within a parallel region due to an uneven distribution among its threads.
$$ OMP_{LB} = (T - T^{I_{serial}} - T^{I_{lb}}) / (T - T^{I_{serial}}) $$

Let $OMP_{Sched}$ ({\em OpenMP Scheduling}) represent the time lost in idle OpenMP threads within a parallel region not caused by load imbalance.
$$ OMP_{Sched} = (T - T^{I_{serial}} - T^{I_{lb}} - T^{I_{sch}}) / (T - T^{I_{serial}} - T^{I_{lb}})$$

Let $OMP_{Eff}$ ({\em OpenMP Parallel Efficiency}) be the efficiency considering only time when OpenMP threads are idle as lost,
$$ OMP_{Eff} = (T - T^I) / T = OMP_{Serial} \times OMP_{LB} \times OMP_{Sched} $$

\paragraph{Redefinition of MPI metrics}
In the hybrid model, the MPI metrics are redefined as:
$$ MPI_{Eff} = (T - T^M) / T $$
$$ MPI_{LB} = (T - T^M) / (max_{i\in[1,N]} T_i - T_{i}^{M}) $$
$$ MPI_{Comm} = (max_{i}^{i\in[1,N]} T_i - T_{i}^{M})/T $$

Please note that, when $M = 1$ (executions with one OpenMP thread or without OpenMP all together), $Hyb_{Eff} = MPI_{Eff}$.
In other words, when $M = 1$, $Hyb_{Eff}$ and all MPI metrics are mathematically equivalent to the ones defined in the base model~\cite{pop_metrics}.

\paragraph{Hybrid parallel efficiency}
Let $Hyb_{Eff}$ ({\em Hybrid Parallel Efficiency}) be the efficiency considering both time in MPI calls and time in idle OpenMP threads as lost:
$$ Hyb_{Eff} = T^U / T $$

\paragraph{Interaction between MPI and OpenMP metrics}
There are certain situations in which threads are idling waiting for an MPI communication to finish.
On the OpenMP side, communication and computation can be overlapped to mitigate this inefficiency.
On the MPI side, reducing the time of the communication would also reduce the time that OpenMP threads are waiting.
Despite both programming models might be at fault,
the current formulation of the hybrid model classifies this situation under the OpenMP metrics.
There is an ongoing work to incorporate interactions between MPI and OpenMP to the hybrid efficiencies model.

\bibliographystyle{unsrt}
\bibliography{999-mn5}

\end{document}